\documentclass[twocolumn,tighten]{aastex631}
\usepackage{graphicx}
\usepackage{amsmath}
\usepackage{natbib}
\usepackage{amssymb}
\usepackage{ulem}
\usepackage{sidecap}

\def\SB{\ensuremath{\sigma_{\rm{SB}}}}

\def\Msun{\ensuremath{\mathrm{M}_\odot}}
\def\Rsun{\ensuremath{\mathrm{R}_\odot}}

\def\tdyn{\ensuremath{t_{\rm dyn}}}
\def\tdiff{\ensuremath{t_{\rm diff}}}
\def\ti{\ensuremath{t_{\rm i}}}
\def\th{\ensuremath{t_{\rm h}}}
\def\tpl{\ensuremath{t_{\rm pl}}}
\def\tbr{\ensuremath{t_{\rm br}}}
\def\Tion{\ensuremath{T_{\rm i}}}

\shorttitle{Maximal Plateau Duration}
\shortauthors{Matsumoto et al.}

\begin{document}

\newcommand{\be}{\begin{equation}}
\newcommand{\ee}{\end{equation}}

\title{Long Plateau Doth So: How Internal Heating Sources Affect Hydrogen-Rich Supernova Light Curves}

\author[0000-0002-9350-6793]{Tatsuya Matsumoto}
\affil{Department of Physics and Columbia Astrophysics Laboratory, Columbia University, Pupin Hall, New York, NY 10027, USA}
\affil{Department of Astronomy, Kyoto University, Kitashirakawa-Oiwake-cho, Sakyo-ku, Kyoto, 606-8502, Japan}
\affil{Hakubi Center, Kyoto University, Yoshida-honmachi, Sakyo-ku, Kyoto, 606-8501, Japan}

\author[0000-0002-4670-7509]{Brian D. Metzger}
\affil{Department of Physics and Columbia Astrophysics Laboratory, Columbia University, Pupin Hall, New York, NY 10027, USA}
\affil{Center for Computational Astrophysics, Flatiron Institute, 162 5th Ave, New York, NY 10010, USA} 

\author[0000-0003-1012-3031]{Jared A. Goldberg}
\affil{Center for Computational Astrophysics, Flatiron Institute, 162 5th Ave, New York, NY 10010, USA}

\begin{abstract}
Some hydrogen-rich core-collapse supernovae (type IIP SNe) exhibit evidence for a sustained energy source powering their light curves, resulting in a brighter and/or longer-lasting hydrogen-recombination plateau phase. We present a semi-analytic SNIIP light curve model that accounts for the effects of an arbitrary internal heating source, considering as special cases $^{56}$Ni/$^{56}$Co decay, a central engine (millisecond magnetar or accreting compact object), and shock interaction with a dense circumstellar disk. While a sustained internal power source can boost the plateau luminosity commensurate with the magnitude of the power, the duration of the recombination plateau can typically be increased by at most a factor $\sim 2-3$ compared to the zero-heating case. For a given ejecta mass and initial kinetic energy, the longest plateau duration is achieved for a constant heating rate at the highest magnitude that does not appreciably accelerate the ejecta. This finding has implications for the minimum ejecta mass required to explain particularly long-lasting supernovae such as iPTF14hls, and for confidently identifying rare explosions of the most-massive hydrogen-rich (e.g. population III) stars.  We present a number of analytic estimates which elucidate the key features of the detailed model.
\end{abstract}


\section{Introduction}
\label{sec:introduction}

Roughly 75\% of all stellar explosions are core-collapse supernovae (SNe; e.g., \citealt{Mannucci+07}), of which $\approx$ 70\% are hydrogen-rich, i.e. type II SNe (e.g., \citealt{Perley+20}).  Roughly 3/4 of the type II SNe are further of the IIP class (hereafter SNIIP), with a plateau-shaped light curve phase lasting typically around 100 days \citep{Barbon+1979,Anderson+2014,Faran+2014,Valenti+2016,Martinez+2022}.  In most cases, the luminosity during the plateau phase is supported mainly by the thermal energy deposited by the supernova shock, its evolution dictated by a cooling and recombination wave receding inwards through the expanding envelope \citep{Grassberg+1971,Grasberg&Nadezhin1976,Falk&Arnett1977,Litvinova&Nadezhin1983,Litvinova&Nadezhin1985,Chugai1991,Popov1993,Faran+2019}.

The properties of SNIIP light curves, such as the duration and luminosity of the plateau phase, can along with spectral information such as photosphere velocities, be used to constrain the ejecta properties (e.g., \citealt{Pejcha&Prieto2015,Martinez+2022b}), though with substantial degeneracies (e.g., \citealt{Dessart&Hillier19},\citealt{Goldberg+2019}). These constraints frequently make use of analytic scaling relations calibrated to numerical simulations (e.g., \citealt{Popov1993,Zha+2023}). It is also known that the sustained heating from the $^{56}$Ni$\rightarrow ^{56}$Co$\rightarrow ^{56}$Fe decay chain acts to flatten and extend the duration of the plateau (e.g., \citealt{Young2004,Kasen&Woosley2009,Bersten+2011,Nakar+2016,Sukhbold+2016,Goldberg+2019,Kozyreva+2019}), which can also be included in numerically-calibrated analytic estimates. 

However, there is increasing evidence for the need for extra energy sources beyond $^{56}$Ni in a growing subset of supernovae, the most extreme cases being superluminous supernovae (SLSNe-II; \citealt{Inserra19,Gal-Yam19}). This ``extra'' heating source has variously been attributed to circumstellar (CSM) interaction (e.g., \citealt{Fassia+00,Smith&McCray07,Miller+09,Inserra+12,Chatzopoulos+2012,Mauerhan+13,Fransson+14,Smith+2014,Bostroem+19,Nyholm+2020}; see \citealt{Fraser20} for a review) or the injection of rotational energy from a rapidly spinning magnetized neutron star \citep{Kasen&Bildsten2010,Woosley2010,Sukhbold&Thompson2017,Dessart2018} or accretion energy from a black hole \citep{Dexter&Kasen2013,Perna+18} or neutron star \citep{Metzger+18b}.  An extreme example is iPTF14hls, which produced a plateau-like (optically-thick) phase lasting for $\sim1000\,\rm days$ \citep{Arcavi+2017d,Sollerman+2019}, only finally to reveal the appearance of narrow emission lines and hence the presence of CSM interaction \citep{Andrews&Smith2018}.   

Extremely luminous or long-lasting SNIIP can also result from rare or exotic explosions of extremely massive progenitors, such as pair-instability supernovae (PISNe) from stars of $\gtrsim 100M_{\odot}$ \citep{Rakavy&Shaviv1967,Heger&Woosley2002,Woosley+2007,Chen+23}. However, the standard analytic expressions commonly used to estimate the ejecta properties from observables like plateau duration and luminosity (e.g., \citealt{Popov1993,Sukhbold+2016}), and hence identify such rare explosion classes, in general do not apply in the presence of additional heating sources.  This motivates the present study to explore the full landscape of theoretically permitted SNIIP properties, allowing for an {\it arbitrary} heating source evolution, in order to ascertain what robust constraints can be placed by supernova observations on the stellar progenitor properties (particularly the ejecta mass). Such a framework would help address whether exceptionally long-duration light curves like iPTF14hls necessitate an atypically massive stellar progenitor, or whether an otherwise ordinary-mass explosion with a powerful internal heating source (e.g., from CSM interaction) is alone sufficient.    

The {\it Vera C.~Rubin Observatory} (\citealt{Ivezic+19}; hereafter {\it Rubin}) will monitor roughly half of the extragalactic sky over ten years, during which it is expected to observe millions of SNIIP \citep{LSST+09}.  This unprecedented sample will enable detailed studies of the distribution of explosion properties (e.g., \citealt{Murphy+19},\citealt{Goldberg&Bildsten2020},\citealt{Martinez+2022}), and how they map onto progenitor star (e.g., \citealt{Strotjohann+2024}) and host galaxy properties (e.g., \citealt{Gagliano+2023}). In addition to the many relatively nearby supernovae observed by {\it Rubin}, rare explosions arising from the first generations of stars (e.g., Population III) are targets for high-redshift surveys such as {\it Euclid} (e.g., \citealt{Moriya+22}) and the {\it Nancy Grace Roman Space Telescope} (e.g., \citealt{Rose+21}). 
The analytic estimates derived in this paper will be usefully applied to constrain the ejecta properties or energy sources for future large samples of SNIIP, or for obtaining robust conclusions for individual exceptional events like iPTF14hls.

This paper is organized as follows. In Sec.~\ref{sec:method} we generalize existing semi-analytic SNIIP light curve models to include the presence of an arbitrary internal heating source.  In Sec.~\ref{sec:constant} we first consider the case of a temporally-constant heating source, providing ample analytic estimates which help interpret our numerical results. In Sec.~\ref{sec:physical} we expand our considerations to the case of a broken-power law heating rate, which we show approximates the behavior of most physical heating sources, ranging from radioactive $^{56}$Ni decay to a central magnetar or accreting engine to CSM interaction. In Sec.~\ref{sec:discussion} we discuss implications of our results and in Sec.~\ref{sec:conclusions} we summarize our findings and conclude. Readers who are not interested in analytical derivations can go directly to Sec.~\ref{sec:discussion} and Fig.~\ref{fig:tL}, where our main findings and their implication are discussed.

\section{Generalized Popov Model}
\label{sec:method}

We generalize the \citet{Popov1993} model to include a central heating source of arbitrary magnitude, which will in essence act to lengthen the optically-thick photospheric phase, i.e. ``plateau'', phase (at least for non-extremal heating rates). While the ``plateau'' should be defined by the phase over which the light curve has a flat shape as identified in observations, defining it in a formal way is not straightforward \citep[see e.g.,][for several definitions in numerical studies]{Sukhbold+2016,Goldberg+2019}. Physically there are two important timescales: One is the diffusion time at which the most photons diffuse out of ejecta. The other is the time when the ejecta becomes optically thin, that is, the end (start) of the photospheric (nebular) phase. During the nebular phase, thermal emission is no longer necessarily produced. As we will see below, under weak energy injection, the light curve drops at the diffusion time and traces the injection luminosity onward. For strong injection, the light curve never experiences a drop and smoothly enters the nebular phase. Since our interest is in obtaining the maximum duration of the plateau under energy injection, we define the plateau phase as the duration over which the ejecta shell is still opaque to electron scattering as the ``plateau'', even when the light curve is not flat. This serves a conservative upper limit on the plateau duration.

Prior to when the ejecta cools sufficiently for hydrogen recombination to set in, its evolution follows the well known Arnett model \citep{Arnett1980,Arnett1982} with an additional heating source \citep[e.g.,][]{Kasen&Bildsten2010,Chatzopoulos+2012,Metzger+2015d}. The first law of thermodynamics applied to a one-zone\footnote{The original models of \cite{Arnett1980,Arnett1982,Popov1993} employ a radial temperature profile corresponding to a solution to the diffusion equation. The simpler one-zone model adopted in this paper nevertheless reproduces the normalization and parameter scalings of these original models, motivating our approach. We also discuss the validation of the one-zone approximation in more detail in Appendix~\ref{asec:validation}.}
\begin{align}
\frac{dE}{dt}&=-P\frac{dV}{dt}-L+H\ ,
	\label{eq:thermodynamics}
\end{align}
where $E$, $P$, $V=\frac{4\pi}{3}R^3$, $R$, $L$, and $H$ are the internal thermal energy, pressure, volume, radius, radiated luminosity, and (specified) heating rate, respectively.  We assume that radiation pressure dominates over gas pressure, such that $E=aT^4V$ and $P=E/(3V)$, respectively, where $a$ is the radiation constant and $T$ is the internal temperature. The radiated luminosity follows from the diffusion approximation: 
\begin{align}
L&\simeq4\pi R^2\frac{c}{3\kappa\rho}\frac{d(aT^4)}{dR}\sim\frac{\tdyn E}{\tdiff ^2}\ ,
    \label{eq:L}
\end{align}
where the dynamical and diffusion timescales are, respectively, defined by 
\begin{align}
\tdyn&\equiv\frac{R}{v}\ ,
    \label{eq:tdyn}\\
\tdiff&\equiv\left(\frac{3\kappa M}{4\pi cv}\right)^{1/2}\simeq110{\,\rm day\,}M_{10}^{1/2}v_{6000}^{-1/2}\ .
    \label{eq:tdiff}
\end{align}
Here $\kappa=0.34\,\rm cm^2\,g^{-1}$ is Thomson opacity for fully ionized solar composition material, $c$ is the speed of light, and we have introduced the shorthand notation $M=10\,M_{10}\,\Msun$ and $v=6000\,v_{6000}\,\rm km\,s^{-1}$ for the ejecta mass and characteristic velocity, respectively. Eq.~\eqref{eq:thermodynamics} can now be written:
\begin{align}
\frac{dE}{dt}=-\frac{E}{\tdyn}-\frac{\tdyn E}{\tdiff^2}+H\ .
    \label{eq:thermodynamics2}
\end{align}

The first term in Eq.~\eqref{eq:thermodynamics} accounts for the loss of internal energy due to $PdV$ work done on the ejecta by itself. This term acts to increase the ejecta kinetic energy $E_{\rm kin}$, according to:
\begin{align}
\frac{dE_{\rm kin}}{dt}=P\frac{dV}{dt}\ .
    \label{eq:EoM}
\end{align}
For homologously expanding ejecta of assumed constant density, $\rho=M/V$, and velocity $v^\prime= v(r/R)$, we have
\begin{align}
E_{\rm kin}&=\int_0^R\left(\frac{1}{2}\rho {v^\prime}^2\right)4\pi r^2dr=\frac{3}{10}Mv^2
    \label{eq:Ekin}\\
&\simeq2.2\times10^{51}{\,\rm erg\,}M_{10}v_{6000}^2\ ,
    \nonumber
\end{align}
Note that we have neglected any contribution to $E_{\rm kin}$ from a high-velocity tail $v' \gg v$, which contributes to the early time shock cooling emission but not to the plateau. Eq.~\eqref{eq:EoM} thus becomes:
\begin{align}
M\frac{dv}{dt}=\frac{20\pi}{3}R^2 P\ .
	\label{eq:EoM2}
\end{align}
Eqs.~\eqref{eq:thermodynamics2} and \eqref{eq:EoM2} determine the time evolution of ejecta until the hydrogen recombination phase begins.

The recombination phase starts when the effective temperature of the photosphere decreases to the hydrogen recombination temperature, $T_{\rm i}\simeq 6000\,\rm K$. The photosphere temperature is defined as:
\begin{align}
T_{\rm eff} \equiv \left(\frac{L}{4\pi\SB R^2}\right)^{1/4} \simeq 1.1\frac{T}{\tau^{1/4}}\ ,
	\label{eq:Teff}
\end{align}
where $\SB$ is the Stefan-Boltzmann constant, $\tau=\kappa R \rho$ is the Thompson optical depth through the ejecta (i.e., assuming it is fully ionized), and the second equality makes use of Eq.~\eqref{eq:L}.

After recombination begins at $t = \ti$, a sharp recombination front forms and begins to recede back through the ejecta shell.  Outside this front, the ejecta is neutral and transparent to radiation.  Therefore we regard the photosphere radius $R_{\rm ph}$ as coinciding with the recombination front, at dimensionless coordinate $x\equiv R_{\rm ph}/R \le 1$.  We retain a one-zone model similar to that described earlier to calculate the evolution of fully ionized region (within the photosphere) during the recombination phase.  All quantities retain the same meaning, except the ejecta radius $R$ is now replaced with $R_{\rm ph} = xR$.  For instance, the volume of the ionized region becomes $V= \frac{4\pi}{3}(xR)^3$, while the radiated luminosity becomes:
\begin{align}
L \simeq 4\pi (xR)^2\frac{c}{3\kappa\rho}\frac{aT^4}{xR}=\frac{\tdyn E}{x^2\tdiff ^2}\ .
    \label{eq:L2}
\end{align}
Equating this to the photosphere luminosity, $L=4\pi (xR)^2\SB \Tion^4$, gives an expression analogous to Eq.~\eqref{eq:Teff} but with replacements $T_{\rm eff}\to T_{\rm i}$ and $\tau\to x\tau$.  Combined with Eq.~\eqref{eq:thermodynamics2} this gives the evolution of the recombination depth \citep{Dexter&Kasen2013}:
\begin{align}
\frac{dx}{dt}&=-\frac{2x}{5\tdyn}-\frac{\tdyn}{5\tdiff^2x}+\frac{H}{5H_{\rm cr}}\frac{1}{\tdyn x^3}\ ,
    \label{eq:dxdt}
\end{align}
where we have defined
\begin{align}
H_{\rm cr}&=4\pi (v\tdiff )^2\SB \Tion^4
    \label{eq:Hcr}\\
&\simeq3.0\times10^{43}{\,\rm erg\,s^{-1}\,}M_{1}v_{6000}T_{\rm i,6000}^4\ ,
    \nonumber
\end{align}
and $\Tion=6000\,T_{\rm i,6000}\,\rm K$. This critical heating rate controls the recombination time and will be used in the analytic estimates below. 

We have neglected acceleration of the ejecta during the recombination phase. This is usually justified because the acceleration timescale at recombination ($t \simeq \ti$),
\begin{align}
\frac{v}{dv/dt}\biggl|_{\ti }=\frac{v}{\frac{5}{4}\kappa aT_{\rm i} ^4}\simeq1700{\,\rm day\,}v_{6000}T_{\rm i,6000}^{-4}\ ,
    \label{eq:condition_ti}
\end{align} 
greatly exceeds the dynamical timescale,\footnote{This may break down for low-velocity events such as supernova precursors \citep[e.g.,][]{Strotjohann+2021,Matsumoto&Metzger2022}. See \cite{Tsuna+2024} for an attempt to include the acceleration during the recombination phase.} where we have used Eqs.~\eqref{eq:EoM2}, \eqref{eq:Teff} and set $T_{\rm eff}=T_{\rm i}$. Acceleration also becomes negligible after radiation begins to diffuse out of the ejecta ($t\gtrsim\tdiff$) due to the associated loss of thermal pressure.

In the following sections, we solve the above equations for different central heating sources $H(t)$ and ejecta properties. The latter includes the total mass $M$, initial size (usually, progenitor star radius) $R_0$, initial internal energy $E_0$, and initial kinetic energy $E_{\rm kin,0}$.\footnote{Radiation pressure dominates over gas pressure for large explosion energies obeying:
\begin{align}
E_0\gtrsim1.3\times10^{48}{\,\rm erg\,}(\mu/0.60)^{-1}M_{10}^{4/3}R_{0,500}^{-1}\ ,
    \label{eq:Prad>Pgas}
\end{align}
where $\mu$ is the mean molecular weight.} Throughout, we assume solar abundances and $\kappa=0.34\,\rm cm^2\,g^{-1}$ (though this could readily be generalized to different abundances by modifying $\mu$ and $\kappa$ accordingly). We also assume that the ejecta internal and kinetic energies are initially comparable, $E_0 \simeq E_{\rm kin,0}$ at $t = 0$, with the total explosion energy equal to their sum. After just a few dynamical times, most of the initial internal energy is converted into kinetic energy by $PdV$ work, causing the ejecta to reach a terminal speed $v=\sqrt{2}v_0$, where $v_0$ is the initial velocity calculated from $E_{\rm kin,0}$ and $M$ (Eq.~\eqref{eq:Ekin}). For $E_{\rm kin,0}=10^{51}\,\rm erg$ and $M = 10M_{\odot}$, this gives $v_0\simeq 4100\,\rm km\,s^{-1}$ and $v\simeq5800\,\rm km\,s^{-1}$. Except where noted (e.g., Sec.~\ref{sec:evolving_v}), we hereafter express analytic estimates in terms of the terminal speed, $v$.

Before describing our results for specific heating sources, we review the original Popov estimates \citep{Popov1993,Dexter&Kasen2013,Matsumoto+2016}, which are obtained neglecting any heating or acceleration of the ejecta after the initial explosion. As shown below (around Eq.~\ref{eq:ti}), the recombination phase begins on the timescale
\begin{align}
\ti&\simeq19{\,\rm day\,}E_{0,51}^{1/2}M_{10}^{-1/2}R_{0,500}^{1/2}v_{6000}^{-1}T_{\rm i,6000}^{-2}\ ,
    \label{eq:ti_popov}
\end{align}
where $E_0=10^{51}E_{0,51}\,\rm erg$, $R_0=500R_{0,500}\,\Rsun$, and we have implicitly assumed $\ti\ll\tdiff$. Under these assumptions, Eq.~\eqref{eq:dxdt} has the following analytic solution:
\begin{align}
x&=\left(\frac{t}{\ti }\right)^{-\frac{2}{5}}\left[1+\frac{1}{7}\left(\frac{\ti }{\tdiff }\right)^{2}-\frac{1}{7}\left(\frac{\ti }{\tdiff }\right)^{2}\left(\frac{t}{\ti }\right)^{\frac{14}{5}}\right]^{\frac{1}{2}}\ .
    \label{eq:x_popov}
\end{align}
The plateau duration corresponds to when the ejecta fully recombines ($x(t_{\rm pl,Popov})=0$), giving:
\begin{align}
t_{\rm pl,Popov}&=\ti \left[1+7\left(\frac{\tdiff }{\ti }\right)^2\right]^{5/14}\simeq7^{5/14}\ti ^{2/7}\tdiff^{5/7}
	\label{eq:tpl_popov}\\
&\simeq130{\,\rm day\,}E_{0,51}^{1/7}M_{10}^{3/14}R_{0,500}^{1/7}v_{6000}^{-9/14}T_{\rm i,6000}^{-4/7}\ ,
    \nonumber
\end{align}
where the second equality on the first line assumes $t_{\rm i}\ll \tdiff$. At the same level of approximation, the total radiated energy and average plateau luminosity are given, respectively, by 
\begin{align}
E_{\rm pl}&=\int_{\ti }^{t_{\rm pl}} Ldt\simeq\frac{56\pi}{55}\SB T_{\rm i}^4v^2\ti^3\left(\frac{t_{\rm i}}{7^{\frac{1}{2}}\tdiff}\right)^{-\frac{11}{7}}\ ,
    \label{eq:Epl_popov}\\
&\simeq2.8\times10^{49}{\,\rm erg\,}E_{0,51}^{5/7}M_{10}^{1/14}R_{0,500}^{5/7}v_{6000}^{-3/14}T_{\rm i,6000}^{8/7}\ ,
    \nonumber\\
L_{\rm pl}&\equiv\frac{E_{\rm pl}}{t_{\rm pl}}\simeq\frac{7^{3/7}56}{55}\pi \SB T_{\rm i}^4v^2t_{\rm i}^{8/7}t_{\rm diff}^{6/7}
    \label{eq:Lpl_popov}\\
&\simeq2.4\times10^{42}{\,\rm erg\,s^{-1}\,}E_{0,51}^{4/7}M_{10}^{-1/7}R_{0,500}^{4/7}v_{6000}^{3/7}T_{\rm i,6000}^{12/7}\ ,
    \nonumber
\end{align}
where we have used $t_{\rm pl,Popov}$ for $t_{\rm pl}$. These expressions exhibit slightly different exponents and normalization from those originally derived by \cite{Popov1993}, who assumed a self-similar radial temperature profile, which is not necessarily achieved \citep[e.g.,][]{Khatami&Kasen2019}. These expressions have been compared with the results of radiation hydrodynamic simulations \citep{Kasen&Woosley2009,Sukhbold+2016,Goldberg+2019}, their accuracy confirmed to within a factor of a few.\footnote{Numerical results in previous works can be summarized as: 
\begin{align}
L_{50}&=\widetilde{L}_{50}{\,\rm erg\,s^{-1}\,}E_{\rm SN,51}^{5/6}M_{10}^{-1/2}R_{0,500}^{2/3}\ ,
    \label{eq:Lpl_ref}\\
t_{\rm pl}&=\widetilde{t}_{\rm pl}{\,\rm day\,}E_{\rm SN,51}^{-1/6}M_{10}^{1/2}R_{0,500}^{1/6}\ ,
    \label{eq:tpl_ref}
\end{align}
where $L_{50}$ is the bolometric luminosity at 50 days after the explosion, and $E_{\rm SN}=10^{51}E_{\rm SN,51}\,\rm erg$ the total explosion energy. \citet{Kasen&Woosley2009} obtained $\widetilde{L}_{50}\simeq1.3\times10^{42}$ and $\widetilde{t}_{\rm pl}\simeq122$ with a different dependence of $t_{\rm pl}\propto E_{\rm SN,51}^{-1/4}$; \citet{Sukhbold+2016} found $\widetilde{L}_{50}\simeq1.8\times10^{42}$ and $\widetilde{t}_{\rm pl}\simeq96$; while \cite{Goldberg+2019} found $\widetilde{L}_{50}\simeq1.4\times10^{42}$ and a slightly different scaling (see their Eq.~8).}

\section{Constant heating rate}
\label{sec:constant}
We now solve the equations presented in the previous section to explore how the supernova plateau phase is modified by a temporally-constant heating source of arbitrary magnitude.  While a constant heating rate only represents a crude approximation to a more physical time-evolving power source, this simple assumption enables us to obtain general physical insights into the impact of heating and to derive useful analytical formulae.  Furthermore, we show later that the constant heating case bounds the light curve behavior for more complex, physically-motivated heating curves, in particular by defining the maximal plateau duration. 

A given model is specified by the (constant) heating rate $H$ and the initial conditions of the explosion, as mentioned above. At early times ($t < \ti$), the effective temperature exceeds the recombination value ($T_{\rm eff}>T_{\rm i}$), and we solve Eq.~\eqref{eq:thermodynamics2} and \eqref{eq:EoM2} until $T_{\rm eff}=T_{\rm i}$ at $\ti$. During the recombination phase, we solve Eq.~\eqref{eq:dxdt} for $x(t)$.  The evolution is followed until the nebular phase, which we define as when the optical depth decreases to unity: $x\tau=1$.

\begin{figure}
\begin{center}
\includegraphics[width=85mm, angle=0, bb=0 0 292 595]{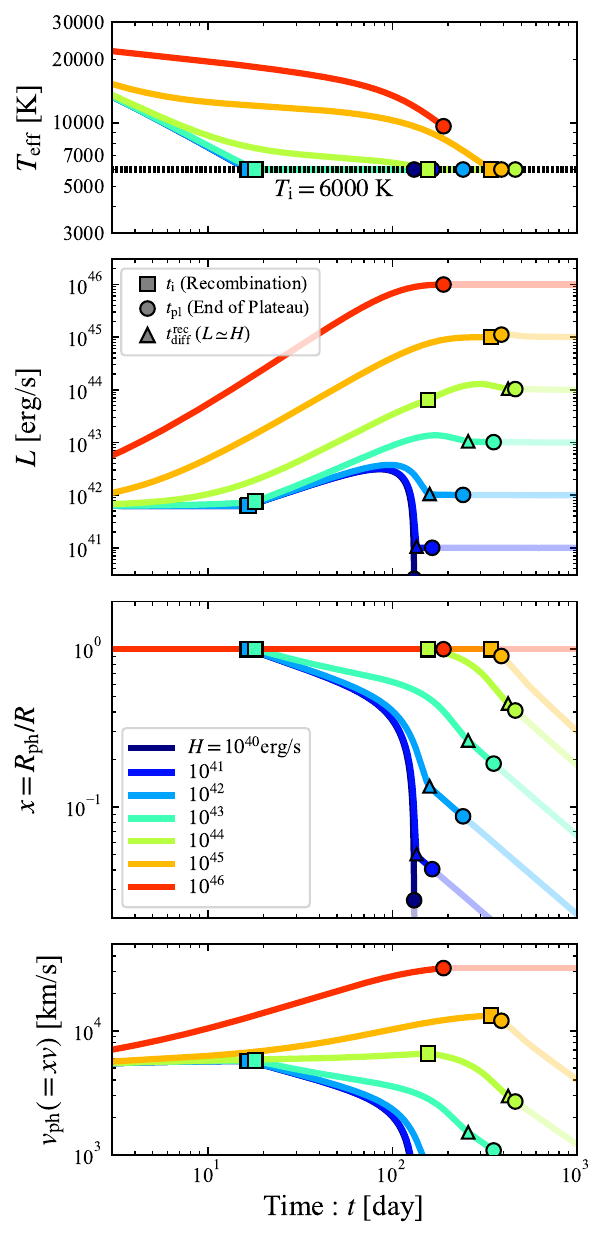}
\caption{Time evolution of the effective temperature, bolometric luminosity, and location and velocity of the photosphere (from top to bottom) for different constant heating rates as marked. The recombination phase begins when $T_{\rm eff}=T_{\rm i}$ at $\ti$ (marked as squares). The ejecta enters the nebular phase when the optical depth becomes smaller than unity at $t_{\rm pl}$ (marked as circles), providing one definition of the plateau duration. 
After time $t_{\rm diff}^{\rm rec}$ (marked as triangles) during the recombination phase, the luminosity starts to track the internal heating rate, a transition which may also be observationally associated with the end of plateau. The adopted ejecta parameters are $M=10\,\Msun$, $R_0=500\,\Rsun$, $E_0=E_{\rm kin,0}=10^{51}\,\rm erg$ ($v_0\simeq4100\,\rm km\,s^{-1}$).}
\label{fig:evolution_h}
\end{center}
\end{figure}

Fig.~\ref{fig:evolution_h} depicts the time evolution of effective temperature, luminosity, and the dimensionless photospheric radius for various values of the heating rate $H = 10^{40}-10^{46}$ erg s$^{-1}$, adopting ejecta properties of $M=10\,\Msun$, $R_0=500\,\Rsun$, and $E_0=E_{\rm kin,0}=10^{51}\,\rm erg$ (corresponding to $v_{0}\simeq4100\,\rm km\,s^{-1}$ and $v\simeq5800\,\rm km\,s^{-1}$), typical of type II SNe \citep[e.g.,][]{Kasen&Woosley2009}. The results are qualitatively summarized as follows: When the heating rate is sufficiently small ($H\lesssim10^{41}\,\rm erg\,s^{-1}$), its effect is negligible and the result is indistinguishable from the zero heating case $H=0$. As the heating rate increases, the plateau duration (equivalently in our formulation, beginning of the nebular phase) becomes longer but the initial onset $\ti$ of the recombination phase remains largely unchanged up to $H\lesssim H_{\rm cr}\simeq3\times10^{43}\,\rm erg\,s^{-1}$. The recombination time grows substantially for $H\gtrsim H_{\rm cr}$. For yet larger heating rates $H \gg H_{\rm cr}$, the recombination phase terminates earlier and the plateau duration shrinks because the ejecta is now significantly accelerated by $PdV$ work. The plateau duration thus achieves its maximal value for a given ejecta mass/explosion energy at the maximum heating rate that does not appreciably impact the ejecta dynamics ($H\simeq10^{44}\,\rm erg\,s^{-1}$, for the adopted parameters).  In the following, we derive analytical formulae for these timescales and the critical heating rates which delineate these regimes. 

\subsection{Without Acceleration}\label{sec:constant_v}
We first consider cases in which $PdV$ acceleration is negligible following the initial dynamical timescale; this greatly simplifies the analysis and is a good approximation for small heating rates. For constant velocity, the dynamical timescale is identical to the time since explosion, $\tdyn=t$ (or equivalently $R=vt$), and Eq.~\eqref{eq:thermodynamics2} can be integrated to obtain: 
\begin{align}
E&=E_0\left(\frac{t_0}{t}\right)e^{-\frac{t^2-t_0^2}{2\tdiff^2}}\left[1+\frac{\tdiff^2H}{t_0E_0}\left(e^{\frac{t^2-t_0^2}{2\tdiff^2}}-1\right)\right]\ ,
    \label{eq:E(t)}
\end{align}
where 
\begin{align}
t_0=\frac{R_0}{v_0}\simeq0.95{\,\rm day\,}R_{0,500}v_{6000}^{-1}\ ,
    \label{eq:t0}
\end{align}
is the initial dynamical/expansion timescale and we again relate the initial and terminal ejecta speed, $v_0=v/\sqrt{2}$. The second term within square brackets in Eq.~\eqref{eq:E(t)} represents the effect of the heating source. For typical supernova ejecta properties, the diffusion time exceeds the initial expansion timescale, $\tdiff\gg t_0$; since initially the heating does not impact the ejecta evolution, the internal energy declines adiabatically, $E\propto t^{-1}$.  This evolution is modified once the deposited energy becomes comparable to the internal energy, $tH\simeq t_0E_0/t$. This occurs on the timescale:
\begin{align}
\th &=\left(\frac{t_0E_0}{H}\right)^{1/2}\simeq33{\,\rm day\,}E_{0,51}^{1/2}R_{0,500}^{1/2}v_{6000}^{-1/2}H_{43}^{-1/2}\ ,
	\label{eq:th}
\end{align}
where $H=10^xH_x\,\rm erg\,s^{-1}$. When this timescale is shorter than the diffusion timescale, which is true for large heating rates 
\begin{align}
H&\gtrsim\frac{t_0E_0}{\tdiff^2}\simeq9.1\times10^{41}{\,\rm erg\,s^{-1}\,}E_{0,51}M_{10}^{-1}R_{0,500}\ ,
    \label{eq:H(th=tdiff)}
\end{align}
the time evolution of the internal energy can be summarized as:
\begin{align}
E&\simeq\begin{cases}
\frac{t_0E_0}{t}&:\,t<\th \ ,\\
tH&:\,\th <t<\tdiff \ ,\\
\frac{\tdiff ^2H}{t}&:\,\tdiff <t\ .\\
\end{cases}
	\label{eq:E_int}
\end{align}
When $\th>\tdiff$, the middle regime disappears.

The effective temperature evolves following Eq.~\eqref{eq:Teff},
\begin{align}
T_{\rm eff}&\simeq\Tion\left(\frac{H}{H_{\rm cr}}\right)^{1/4}\begin{cases}
\left(\frac{t}{t_{\rm h}}\right)^{-1/2}&:\,t<\th \ ,\\
1&:\,\th <t<\tdiff \ ,\\
\left(\frac{t}{\tdiff}\right)^{-1/2}&:\,\tdiff <t\ ,\\
\end{cases}
	\label{eq:Teff1_h}
\end{align}
where $H_{\rm cr}$ is the critical heating rate introduced earlier (Eq.~\eqref{eq:Hcr}). The recombination time $\ti$ is obtained by setting $T_{\rm eff}=T_{\rm i}$ with Eq.~\eqref{eq:Teff1_h}. Noting that the effective temperature is constant for $\th<t<\tdiff$, we obtain 
\begin{align}
\ti&=\begin{cases}
\left(\frac{t_0E_0}{H_{\rm cr}}\right)^{1/2}&:\,H\leq H_{\rm cr}\ ,\\
\left(\frac{\tdiff^2H}{H_{\rm cr}}\right)^{1/2}&:\,H_{\rm cr}< H\ ,\\
\end{cases}
	\label{eq:ti}\\
&\simeq\begin{cases}
19{\,\rm day\,}E_{0,51}^{1/2}M_{10}^{-1/2}R_{0,500}^{1/2}v_{6000}^{-1}T_{\rm i,6000}^{-2}&:\,H\leq H_{\rm cr}\ ,\\
66{\,\rm day\,}v_{6000}^{-1}T_{\rm i,6000}^{-2}H_{43}^{1/2}&:\,H_{\rm cr}< H\ .
\end{cases}
    \nonumber
\end{align}
This timescale for $H<H_{\rm cr}$ is notably identical to that in the $H=0$ limit (Eq.~\eqref{eq:ti_popov}). Interestingly, the recombination time is discontinuous at $H=H_{\rm cr}$ because of the static effective temperature for $\th<t<\tdiff$, thus accounting for the sudden jump in the recombination time above $H \simeq H_{\rm cr} \simeq 3\times10^{43}\,\rm erg\,s^{-1}$.

For the recombination phase, while Eq.~\eqref{eq:dxdt} cannot be solved analytically for $H\neq0$, the behavior of the solution can still be understood quantitatively. The three terms on the right hand side of Eq.~\eqref{eq:dxdt} correspond to adiabatic cooling, radiative cooling (photon diffusion), and heating, respectively. First consider the limit of weak heating $H<H_{\rm cr}$. In such cases, just after the recombination starts, neither heating or radiative losses are important, and the photosphere thus shrinks in time as $x\simeq(t/\ti)^{-2/5}$ corresponding to adiabatic evolution (Eq.~\eqref{eq:x_popov}). However, as $x$ decreases, radiative cooling and heating eventually come to compete with, or balance, respectively, the adiabatic cooling.  For parameter regimes in which radiative cooling is subdominant, adiabatic cooling balances heating. This balance is achieved on the timescale 
\begin{align}
t_{\rm h}^{\rm rec}&=\left(\frac{2H_{\rm cr}}{H}\right)^{5/8}\ti
	\label{eq:th^rec}\\
&\simeq59{\,\rm day\,}E_{0,51}^{1/2}M_{10}^{1/8}R_{0,500}^{1/2}v_{6000}^{-3/8}T_{\rm i,6000}^{1/2}H_{43}^{-5/8}\ .
	\nonumber
\end{align}
At times $t > t_{\rm h}^{\rm rec}$, the photosphere stalls at $x \simeq (H/2H_{\rm cr})^{1/4}$ until radiative cooling takes over from adiabatic cooling in balancing heating.  This occurs after on the timescale:
\begin{align}
t_{\rm diff}^{\rm rec}&=\left(\frac{2H}{H_{\rm cr}}\right)^{1/4}\tdiff\simeq99{\,\rm day\,}M_{10}^{1/4}v_{6000}^{-3/4}T_{\rm i,6000}^{-2}H_{43}^{1/4}\ .
	\label{eq:tdiff^rec}
\end{align}
After this transition $t > t_{\rm diff}^{\rm rec}$, the photosphere continues to shrink again, this time as $x\simeq (H/H_{\rm cr})^{1/2}\tdiff/t$. In summary, for $H<H_{\rm cr}$ and $\ti<\th<\tdiff$, the photosphere location evolves as:
\begin{align}
x&\simeq\begin{cases}
\left(\frac{t}{\ti }\right)^{-2/5}&:\,t<t_{\rm h}^{\rm rec} \ ,\\
\left(\frac{H}{2H_{\rm cr}}\right)^{1/4}&:\,t_{\rm h}^{\rm rec} <t<t_{\rm diff}^{\rm rec} \ ,\\
\left(\frac{H}{H_{\rm cr}}\right)^{1/2}\frac{\tdiff}{t}&:\,t_{\rm diff}^{\rm rec} <t\ .
\end{cases}
	\label{eq:x(t)}
\end{align}
The superscripts ``rec" in Eqs.~\eqref{eq:th^rec} and \eqref{eq:tdiff^rec} denote heating and diffusion timescales during the recombination phase, respectively. After the diffusion timescale, the escaping luminosity roughly tracks the heating rate $L\simeq H$, as illustrated by triangles in Fig.~\ref{fig:evolution_h}. While we have defined the end of plateau as when the ejecta becomes optically thin, one could instead define this by the change in light curve shape near $t_{\rm diff}^{\rm rec}$ (e.g. the location of the triangles for the $H=10^{41}$ and $10^{42}\,\rm erg\,s^{-1}$ cases in Fig.~\ref{fig:evolution_h}). We discuss later in this section how our results changes if we instead identify $t_{\rm diff}^{\rm rec}$ as the plateau duration.

For the opposite regime, radiative cooling dominates adiabatic cooling even before either balances heating. In these cases the photosphere evolution follows Eq.~\eqref{eq:x_popov} until heating balances cooling at $t \simeq t_{\rm pl,Popov}$, after which $x$ decreases following the final case in Eq.~\eqref{eq:x(t)}. Such an evolutionary sequence is realized for low heating rates, below a critical value:
\begin{align}
H&\lesssim 0.66\left(\frac{\ti}{\tdiff}\right)^{8/7}H_{\rm cr}\\
&\simeq2.7\times10^{42}{\,\rm erg\,s^{-1}\,}E_{0,51}^{4/7}M_{10}^{-1/7}R_{0,500}^{4/7}v_{6000}^{3/7}T_{\rm i,6000}^{12/7}\ ,
	\nonumber
\end{align}
obtained by equating the plateau duration neglecting heating ($H=0$; Eq.~\eqref{eq:tpl_popov}) with the timescale \eqref{eq:th^rec}. For large heating rates $H>H_{\rm cr}$, the first regime in Eq.~\eqref{eq:x(t)} vanishes because heating has already become important even prior to the recombination phase.

In either case, as long as the heating is present, the photosphere at late times $t> t_{\rm diff}^{\rm rec}$ or $t_{\rm pl,Popov}$ begins to shrink as $x\propto t^{-1}$ (following the final regime of Eq.~\eqref{eq:x(t)}) and never reaches $x=0$ as in the no-heating case $H=0$. Again defining the plateau duration at when the ejecta becomes optically thin (or enters the nebular phase): $x\tau=1$, in the presence of a heating source we find:
\begin{align}
t_{\rm pl}&\simeq \left(\frac{v}{c}\right)^{1/6}\left(\frac{H}{H_{\rm cr}}\right)^{1/6}t_{\rm thin}=\left(\frac{3^2\kappa^2M^2H}{2^6\pi^3\SB T_{\rm i}^4v^6}\right)^{1/6}
    \label{eq:tpl}\\
&\simeq340{\,\rm day\,}M_{10}^{1/3}v_{6000}^{-1}T_{\rm i,6000}^{-2/3}H_{43}^{1/6}\ ,
    \nonumber
\end{align}
where 
\begin{align}
t_{\rm thin}&=\left(\frac{3\kappa M}{4\pi v^2}\right)^{1/2}\simeq780{\,\rm day\,}M_{10}^{1/2}v_{6000}^{-1}\ ,
	\label{eq:tthin}
\end{align}
is the time the ejecta becomes optically thin assuming full ionization.  Equating Eq.~\eqref{eq:tpl} with Eq.~\eqref{eq:tpl_popov} gives the heating rate above which the plateau duration is significantly boosted relative to the zero-heating case:
\begin{align}
H_{\rm min}&=\left(\frac{c}{v}\right)\left(\frac{t_{\rm pl,Popov}}{t_{\rm thin}}\right)^{6}H_{\rm cr}
	\label{eq:Hmin}\\
&\simeq3.9\times10^{40}{\,\rm erg\,s^{-1}\,}E_{0,51}^{6/7}M_{10}^{-5/7}R_{0,500}^{6/7}v_{6000}^{15/7}T_{\rm i,6000}^{4/7}\ .
	\nonumber
\end{align}
Note that while heating can extend the plateau duration in accordance with Eq.~\eqref{eq:tpl}, the duration is ultimately limited by when the ejecta becomes optically-thin/nebular (Eq.~\eqref{eq:tthin}). The latter limit is achieved for very powerful heating sources, which keep the ejecta fully ionized to late times. However, as we shall discuss, sufficiently powerful energy sources also act to accelerate the ejecta, hastening the nebular phase (Eq.~\eqref{eq:tthin}).

\subsection{With Acceleration}\label{sec:evolving_v}
We now consider the effects of ejecta acceleration, which modifies the characteristic timescales ($\tdiff$ and $\tpl$) which depend on $v$.  Because the internal energy evolution is not modified from Eq.~\eqref{eq:E_int} until the photon diffusion time, Eq.~\eqref{eq:EoM2} can be rewritten using the approximation $R\simeq vt$:
\begin{align}
\frac{dv}{dt}=\frac{5E}{3MR}\simeq\begin{cases}
\frac{5t_0E_0}{3Mvt^2}&:\,t<\th\ ,\\
\frac{5H}{3Mv}&:\,\th<t<t_{\rm diff}^{\rm acc}\ .
\end{cases}
    \label{eq:EoM3}
\end{align}
Here $t_{\rm diff}^{\rm acc}$ is the diffusion timescale, now including the effects of acceleration, which in general will differ from Eq.~\eqref{eq:tdiff} as described below. Eq.~\eqref{eq:EoM3} can be integrated to obtain: 
\begin{align}
v&\simeq\begin{cases}
\sqrt{v_0^2+v_0^2\left(1-\frac{t_0}{t}\right)}&:\,t<\th\ ,\\
\sqrt{v_0^2+v_0^2\left(1-\frac{t_0}{\th}\right)+\frac{10H(t-\th)}{3M}}&:\,\th<t<t_{\rm diff}^{\rm acc}\ ,
\end{cases}
    \label{eq:v(t)}
\end{align}
where $v_0$ is the initial ejecta velocity. As discussed earlier, for $t<\th$, the velocity increases and asymptotes to $v = \sqrt{2}v_0$ at the expense of the initial internal energy, regardless of any heating source. For $t>\th$, the acceleration by $PdV$ work becomes significant on the timescale
\begin{align}
t_{\rm acc}=\frac{3Mv_0^2}{5H}\simeq22{\,\rm day\,}M_{10}v_{0,4000}^2H_{45}^{-1}\ ,
    \label{eq:tacc}
\end{align}
as determined by the condition that the injected energy be comparable to the ejecta kinetic energy ($t_{\rm acc}H \simeq E_{\rm kin}$). Here $v_0=4000v_{0,4000}\,\rm\,km\,s^{-1}$. Therefore, the velocity evolution until $t<t_{\rm diff}^{\rm acc}$ can be summarized as: 
\begin{align}
v&\simeq\sqrt{2}v_0\begin{cases}
1&:\,t<t_{\rm acc}\ ,\\
\left(\frac{t}{t_{\rm acc}}\right)^{1/2}&:\,t_{\rm acc}<t<t_{\rm diff}^{\rm acc}\ .
\end{cases}
	\label{eq:v(t)2}
\end{align}

The diffusion timescale is obtained by substituting Eq.~\eqref{eq:v(t)2} into Eq.~\eqref{eq:tdiff} and setting $t=t_{\rm diff}^{\rm acc}$:
\begin{align}
t_{\rm diff}^{\rm acc}=\left(\frac{27\kappa^2M^3}{160\pi^2c^2H}\right)^{1/5}\simeq82{\,\rm day\,}M_{10}^{3/5}H_{45}^{-1/5}\ .
	\label{eq:tdiff^acc}
\end{align}
We find that after radiative cooling becomes important for $t>t_{\rm diff}^{\rm acc}$, the internal energy starts to decline again, following Eq.~\eqref{eq:E_int}.  The equation of motion then takes the form $dv/dt\propto1/(vt)^2$ and gives a similar solution to Eq.~\eqref{eq:v(t)} for $t<\th$. We can thus regard that acceleration effectively ceases at $t_{\rm diff}^{\rm acc}$. The maximal velocity attained for $\th<t<t_{\rm diff}^{\rm acc}$ is given by Eqs.~\eqref{eq:v(t)2} and \eqref{eq:tdiff^acc}:
\begin{align}
v_{\rm max}\simeq\left(\frac{25\kappa H^2}{3\pi cM}\right)^{1/5}\simeq11000{\,\rm km\,s^{-1}\,}M_{10}^{-1/5}H_{45}^{2/5}\ .
    \label{eq:v2}
\end{align}
The corresponding effective temperature evolution is obtained by substituting Eqs.~\eqref{eq:v(t)2} or \eqref{eq:v2} into Eq.~\eqref{eq:Teff1_h}:
\begin{align}
T_{\rm eff}&=\begin{cases}
\left(\frac{c^2H}{30\kappa^2\SB^2Mt}\right)^{1/8}&:\,t_{\rm acc} <t<t_{\rm diff}^{\rm acc}\ ,\\
\left(\frac{9c^2M^2H}{2^{10}5^4\pi^3\SB^5\kappa^2t^{10}}\right)^{1/20}&:\,t_{\rm diff}^{\rm acc} <t\ .\\
\end{cases}
    \label{eq:Teff2_h}
\end{align}
During the acceleration phase, the effective temperature remains almost constant and recombination starts after $t_{\rm diff}^{\rm acc}$, as in the constant velocity case.\footnote{Recombination occurs during the acceleration phase ($t_{\rm acc}<t<t_{\rm diff}^{\rm acc}$) for
\begin{align}
H\lesssim\left(\frac{3^45^2\SB^5\kappa^6M^4T_{\rm i}^{20}}{\pi c^6}\right)^{1/3}\simeq7.8\times10^{42}{\,\rm erg\,s^{-1}\,}M_{10}^{4/3}T_{\rm i,6000}^{20/3}\ .
\nonumber
\end{align}
Such small heating rates do not significantly impact the ejecta's dynamics.} 
This timescale is obtained by equating Eq.~\eqref{eq:Teff2_h} with $T_{\rm i}$:
\begin{align}
t_{\rm i}^{\rm acc}&=\left(\frac{9c^2M^2H}{2^{10}5^4\pi^3 \SB^5\kappa^2T_{\rm i}^{20}}\right)^{1/10}
    \label{eq:ti^acc}\\
&\simeq350{\,\rm day\,}M_{10}^{1/5}T_{\rm i,6000}^{-1/2}H_{45}^{1/10}\ .
    \nonumber
\end{align}

Once recombination begins at $t_{\rm i}^{\rm acc}$, the following evolution is identical to the constant velocity case. In particular, the plateau duration is obtained by substituting the maximal velocity, Eq.~\eqref{eq:v2}, into Eq.~\eqref{eq:tpl}:
\begin{align}
t_{\rm pl}^{\rm acc}&=\left(\frac{3^{16}\kappa^4c^6M^{16}}{2^{30}5^{12}\pi^9\SB^5T_{\rm i}^{20}H^7}\right)^{1/30}
    \label{eq:tpl^acc}\\
&\simeq400{\,\rm day\,}M_{10}^{8/15}T_{\rm i,6000}^{-2/3}H_{45}^{-7/30}\ ,
    \nonumber
\end{align}
Contrary to the constant-velocity case, the plateau duration now shrinks with increasing $H$ as a result of the higher ejecta expansion speed.

{\it The maximal plateau duration is thus obtained by the maximal heating rate which does not appreciably affect the ejecta dynamics.}  Since acceleration of the ejecta ceases on the diffusion time, we can obtain this critical heating rate by equating $t_{\rm acc}$ with $t_{\rm diff}^{\rm acc}$ (Eq.~\eqref{eq:tdiff^acc}):
\begin{align}
H_{\rm pl,max}&=\left(\frac{3\pi cMv^5}{25\kappa}\right)^{1/2}
    \label{eq:Hpl_max}\\
&\simeq2.3\times10^{44}{\,\rm erg\,s^{-1}\,}M_{10}^{1/2}v_{6000}^{5/2}\ ,
	\nonumber
\end{align}
which when substituted into Eq.~\eqref{eq:tpl} gives the corresponding {\it maximal} plateau duration:
\begin{align}
t_{\rm pl,max}&=\left(\frac{3^5\kappa^3c M^5}{2^{12}5^2\pi^5\SB^2T_{\rm i}^8v^7}\right)^{1/12}
    \label{eq:tpl_max}\\
&\simeq570{\,\rm day\,}M_{10}^{5/12}v_{6000}^{-7/12}T_{\rm i,6000}^{-2/3}\ .
    \nonumber
\end{align}

Above we have assumed that the acceleration terminates on the diffusion time assuming fully ionized ejecta; however, this transition can occur earlier if the ejecta recombines first. Because recombination occurs when $T_{\rm eff}=T_{\rm ion}$, the maximal heating rate in this case is $H_{\rm cr}$ (Eq.~\eqref{eq:Hcr}). When the condition 
\begin{align}
E_{\rm kin}<H_{\rm cr}\tdiff\ ,
    \label{eq:condition_tpl}
\end{align}
is satisfied, the critical heating rate obeys $H_{\rm cr}>H_{\rm pl,max}$, resulting in a maximal plateau duration:
\begin{align}
t_{\rm pl,max}&=\left(\frac{3^3\kappa^3M^3}{2^6\pi^3cv^5}\right)^{1/6}\simeq410{\,\rm day\,}M_{10}^{1/2}v_{6000}^{-5/6}\ .
    \label{eq:tpl_max2}
\end{align}

For very large heating rates, the ejecta bypasses the recombination phase entirely and proceeds to become optically thin while still fully ionized. By equating Eq.~\eqref{eq:tpl^acc} with Eq.~\eqref{eq:ti^acc}, we obtain the critical heating rate above which the recombination phase vanishes:
\begin{align}
H_{\rm thin}&=\left(\frac{c}{v}\right)H_{\rm cr}=3\kappa\SB \Tion^4M
    \label{eq:Hthin}\\
&\simeq1.5\times10^{45}{\,\rm erg\,s^{-1}\,}M_{10}T_{\rm i,6000}^4\ .
    \nonumber
\end{align}
The nebular phase in this case begins at 
\begin{align}
t_{\rm thin}^{\rm acc}=\left(\frac{3^7\kappa^3c^2M^7}{2^{10}5^4\pi^3H^4}\right)^{1/10}\simeq430{\,\rm day\,}M_{10}^{7/10}H_{45}^{-2/5}\ ,
	\label{eq:tthin^acc}
\end{align}
obtained by substituting Eq.~\eqref{eq:v2} into Eq.~\eqref{eq:tthin}.

\begin{figure}
\begin{center}
\includegraphics[width=85mm, angle=0, bb=0 0 283 228]{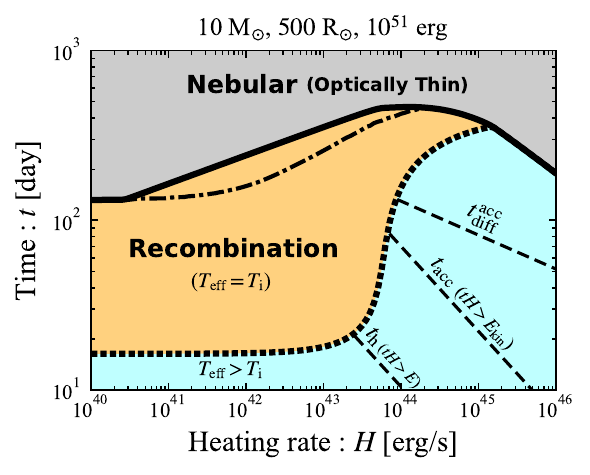}
\caption{Key timescales as a function of the assumed constant ejecta heating rate. Dotted and solid curves denote the recombination time $\ti$ and plateau duration $\tpl$, respectively. Light blue, orange, and gray shaded regions represent the photospheric phase with $T_{\rm eff}>\Tion$, recombination phase ($T_{\rm eff}=\Tion$), and nebular phase (optically thin), respectively. The dash-dotted curve shows $t_{\rm diff}^{\rm rec}$, after which the luminosity begins to track the input heating rate. Dashed lines represent three characteristic timescales, on which the heating appreciably alters the thermal ($t_{\rm h}$, Eq.~\eqref{eq:th}) and dynamical ($t_{\rm acc}$, Eq.~\eqref{eq:tacc}) evolution of the ejecta, as well as the diffusion time for maximally-accelerated ejecta ($t_{\rm diff}^{\rm acc}$, Eq.~\eqref{eq:tdiff^acc}).}
\label{fig:time_h}
\end{center}
\end{figure}

Fig.~\ref{fig:time_h} illustrates the key timescales as a function of the heating rate. Black dotted and solid curves show the recombination time $\ti$ and plateau duration $\tpl$, respectively, as obtained by solving the equations directly as in Fig.~\ref{fig:evolution_h}. Their behavior agrees with our analytical formulae, supporting the above arguments. A dotted curve denotes the recombination time $\ti$ at which the effective temperature equals the recombination temperature. For small heating rates $H < H_{\rm cr}\simeq3\times10^{43}{\,\rm erg\,s^{-1}}$ (Eq.~\eqref{eq:Hcr}), this timescale is independent of $H$ and identical to the Popov formula (Eqs.~\eqref{eq:ti_popov}, \eqref{eq:ti}). For heating rates approaching the critical value, $H\simeq H_{\rm cr}$, the recombination time increases rapidly (representing the gap in $\ti$ in Eq.~\eqref{eq:ti}), because heating begins to affect the thermal evolution for $t\gtrsim t_{\rm h}$ (Eq.~\eqref{eq:th}), which significantly delays the recombination.  Once heating starts to affect the dynamics of the ejecta as well for $t_{\rm acc}<t<t_{\rm diff}^{\rm acc}$ (Eqs.~\ref{eq:tacc} and \ref{eq:tdiff^acc}), the recombination time increases more gradually, as $t_{\rm i}^{\rm acc}\propto H^{1/10}$ (Eq.~\eqref{eq:ti^acc}).

A solid curve in Fig.~\ref{fig:time_h} depicts the plateau time $\tpl$ at which the ejecta becomes optically thin and enters the nebular phase.  For heating rates $H\gtrsim H_{\rm min}\simeq4\times10^{40}{\,\rm erg\,s^{-1}}$ (Eq.~\eqref{eq:Hmin}), the plateau duration is prolonged from the Popov formula (Eq.~\eqref{eq:tpl_popov}) as $t_{\rm pl}\propto H^{1/6}$ (Eq.~\eqref{eq:ti}), where $H_{\rm min}$ is smaller than the critical rate $H_{\rm cr}$ because the heating plays a role only during the recombination phase.  When the acceleration becomes significant for higher heating rates, the plateau duration shrinks as $t_{\rm pl}^{\rm acc}\propto H^{-7/30}$ (Eq.~\eqref{eq:tpl^acc}) upon further heating due to the higher ejecta velocity. The maximal allowed plateau duration for a given ejecta mass (Eq.~\eqref{eq:tpl_max}) is thus obtained for the maximal heating rate (Eq.~\eqref{eq:Hpl_max}) which does not appreciably accelerate the ejecta.  For the highest heating rates $H>H_{\rm thin}\simeq2\times10^{45}{\,\rm erg\,s^{-1}}$ (Eq.~\eqref{eq:Hthin}), the ejecta remains ionized throughout its entire evolution and the recombination phase disappears. The nebular phase also begins earlier for largest hearing rate as $t_{\rm thin}^{\rm acc}\propto H^{-2/5}$ (Eq.~\eqref{eq:tthin^acc}) because of the higher velocity.

As discussed around Eq.~\eqref{eq:tdiff^acc}, the plateau duration can observationally be identified as when the luminosity drops rapidly and begins to track the ejecta heating rate \citep[e.g.,][]{Valenti+2016}. In our framework, this transition happens at $t_{\rm diff}^{\rm rec}$, as indicated by a dash-dotted curve in Fig.~\ref{fig:time_h}. By definition this transition occurs before $t_{\rm pl}$ and thus provides a conservative lower limit on the plateau duration, while it does not change our quantitative results significantly.

\begin{figure}
\begin{center}
\includegraphics[width=85mm, angle=0, bb=0 0 285 225]{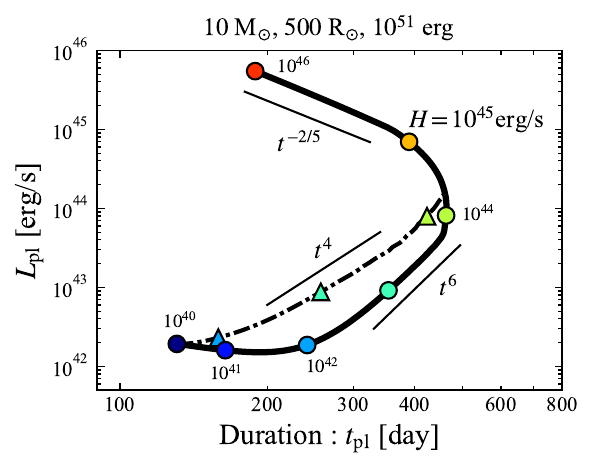}
\includegraphics[width=85mm, angle=0, bb=0 0 287 213]{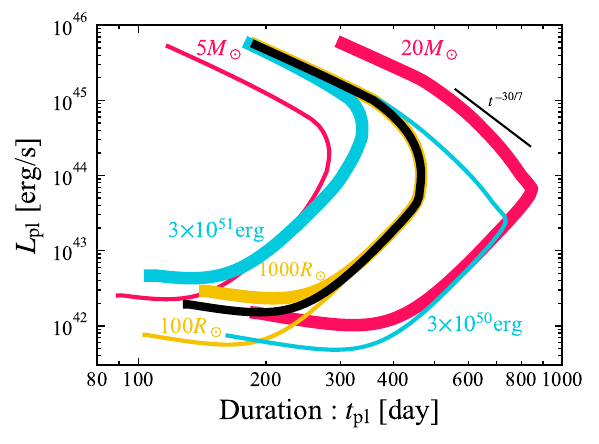}
\caption{({\bf Top}) Trajectory of plateau duration and luminosity for different constant heating rates (increasing in order-of-magnitude increments from blue to red) for an canonical explosion of ejecta mass $10M_{\odot}$, energy $10^{51}$ erg, and 500$R_{\odot}$ progenitor radius. Circle and triangles along the curve shows the results corresponding to calculations in Fig.~\ref{fig:evolution_h} with the same color scheme. ({\bf Bottom}) The same as the top panel, but for different ejecta properties (total mass, initial energy, and progenitor radius) as marked.}
\label{fig:tL_h}
\end{center}
\end{figure}

Fig.~\ref{fig:tL_h} shows the ``trajectory'' of plateau duration and luminosity, achieved for different heating rates, in the duration-luminosity phase space diagram of optical transients \citep{Kasliwal2011}. As with the Popov formulae (Eq.~\eqref{eq:Epl_popov}), we define the plateau luminosity by dividing the total radiated energy up to the nebular phase by the plateau duration (when the ejecta become optically thin):
\begin{align}
L_{\rm pl}\equiv \frac{E_{\rm rad}}{t_{\rm pl}}\,\,\,\text{and}\,\,\,E_{\rm rad}=\int_0^{t_{\rm pl}}Ldt\ .
\end{align}
For $H\lesssim H_{\rm thin}$, the total radiated energy is dominated by the recombination phase. Initially, as the heating rate increases, only the plateau duration increases (Fig.~\ref{fig:evolution_h}, Eq.~\eqref{eq:tpl}). This incremental change reflects as only a modest flattening of the light curve after the main plateau for $t_{\rm diff}^{\rm rec}<t<t_{\rm pl}$, which does not contribute to the radiated energy, causing the plateau luminosity to remain unchanged or to slightly decrease. As $H$ increases further, both the luminosity and duration increase up to the critical heating rate which gives the maximal plateau duration. In this regime, the radiated energy is dominated by the recombination phase:
\begin{align}
E_{\rm rad}&\simeq\int_{\ti}^{t_{\rm pl}}Ldt=4\pi v^2\SB T_{\rm i}^4\int_{\ti}^{t_{\rm pl}}(xt)^2dt\simeq Ht_{\rm pl}\ ,
\end{align}
where the last equality makes use of $x\simeq(H/H_{\rm cr})^{1/2}t_{\rm diff}/t$ (Eq.~\eqref{eq:x(t)}) and assumes $t_{\rm pl}\gg t_{\rm i}$. Perhaps surprisingly, this expression also provides a reasonable estimate of the the plateau luminosity even for larger heating rates, because the supernova luminosity closely tracks $H$ after the diffusion timescale (see Fig.~\ref{fig:evolution_h}). By solving Eq.~\eqref{eq:tpl}, \eqref{eq:tpl^acc}, and \eqref{eq:tthin^acc} for $H$, we obtain the following scaling relations:
\begin{align}
L_{\rm pl}&\simeq\begin{cases}
3.9\times10^{42}{\,\rm erg\,s^{-1}\,}E_{\rm kin,0,51}^3M_{10}^{-5}T_{\rm i,6000}^4\left(\frac{t_{\rm pl}}{300{\,\rm day}}\right)^{6}\\
\,\,\,\,\,\,\,\,\,\,\,\,\,\,\,\,\,\,\,\,\,\,\,\,\,\,\,\,\,\,\,\,\,\,\,\,\,\,\,\,\,\,\,\,\,\,\,\,\,\,\,\,\,\,\,\,\,\,\,\,\,\,\,\,\,\,\,\,\,\,\,\,:H<H_{\rm pl,max}\ ,\\
3.5\times10^{45}{\,\rm erg\,s^{-1}\,}M_{10}^{16/7}T_{\rm i,6000}^{-20/7}\left(\frac{t_{\rm pl}}{300{\,\rm day}}\right)^{-30/7}\\
\,\,\,\,\,\,\,\,\,\,\,\,\,\,\,\,\,\,\,\,\,\,\,\,\,\,\,\,\,\,\,\,\,\,\,\,\,\,\,\,\,\,\,\,\,\,\,\,\,\,\,\,\,\,\,\,\,\,\,\,\,\,\,\,\,\,\,\,\,\,\,\,:H_{\rm pl,max}<H<H_{\rm thin}\ ,\\
2.5\times10^{45}{\,\rm erg\,s^{-1}\,}M_{10}^{7/4}\left(\frac{t_{\rm pl}}{300{\,\rm day}}\right)^{-5/2}\\
\,\,\,\,\,\,\,\,\,\,\,\,\,\,\,\,\,\,\,\,\,\,\,\,\,\,\,\,\,\,\,\,\,\,\,\,\,\,\,\,\,\,\,\,\,\,\,\,\,\,\,\,\,\,\,\,\,\,\,\,\,\,\,\,\,\,\,\,\,\,\,\,:H_{\rm thin}<H\ ,
\end{cases}
	\label{eq:Lpl_tpl}
\end{align}
where we have again replaced $t_{\rm pl}^{\rm acc}$ in the middle and $t_{\rm thin}^{\rm acc}$ in the bottom regimes with $t_{\rm pl}$, retaining the same notation. We have also substitute $v$ from Eq.~\eqref{eq:Ekin}.

The top panel of Fig.~\ref{fig:tL_h} depicts the $L_{\rm pl}-t_{\rm pl}$ trajectories of our light curve models from Fig.~\ref{fig:evolution_h}, illustrating how our analytic scaling relations (Eq.~\eqref{eq:Lpl_tpl}) can reasonably reproduce the results of our full numerical calculations. For the specific ejecta properties chosen, the middle regime of Eq.~\eqref{eq:Lpl_tpl} does not clearly appear, because the two characteristic heating rates, $H_{\rm pl,max}\sim H_{\rm thin}$, are not well separated (though this regime appears more clearly for other parameter choices; see below). A dash-dotted curve shows how the trajectory changes if one were to instead take $t_{\rm diff}^{\rm rec}$ (the time at which the light curve starts to track the central heating source) as the definition of plateau duration. This alternative definition predicts a slightly different scaling relationship:
\begin{align}
L_{\rm pl}\simeq2.3\times10^{43}{\,\rm erg\,s^{-1}\,}E_{\rm kin,0,51}^{3/2}M_{10}^{-5/2}T_{\rm i,6000}^4\left(\frac{t_{\rm pl}}{300{\,\rm day}}\right)^4\ .
    \nonumber
\end{align}

The bottom panel of Fig.~\ref{fig:tL_h} shows similar $L_{\rm pl}-t_{\rm pl}$ trajectories, but for different ejecta and explosion properties. While the shape of the trajectory does not change significantly (except for the middle regime in Eq.~\eqref{eq:Lpl_tpl} appearing in some cases), it shifts side to side, depending on the kinetic energy and ejecta mass. These general behavior can be simply understood by the longer diffusion timescale/plateau duration that arises for more massive and/or less-energetic/slower ejecta. Consistent with Eq.~\eqref{eq:Lpl_tpl}, the trajectory depends only weakly on the initial progenitor radius $R_0$, except in the weakly heated regime $H<H_{\rm min}$ corresponding to the standard Popov limit.

\section{Physically motivated heating sources}
\label{sec:physical}

We now apply our model to several physically-motivated ejecta heating sources: radioactive $^{56}$Ni/$^{56}$Co decay; rotational or accretion power from a central compact object (neutron star or black hole); and shock interaction with circumstellar gas. Some of these heating sources may be spatially localized within the ejecta, rendering our one-zone assumption questionable. However, we believe our model still provides useful physical insight and a reasonable approximation for the impact on the plateau duration in these cases, particularly in light of the uncertainties that impact the exact form of the heating in many of these cases. Indeed, our results for magnetar powered light curves match well results obtained by radiation hydrodynamic simulations, which make the extreme assumption of a completely centralized heating source (see Sec.~\ref{sec:magnetar}). We also give a more detailed justification of the one-zone model in Appendix~\ref{asec:validation}.

The evolution of the heating rate can in most of these cases be approximately described as a constant heating rate up to some break time $t_{\rm br}$, followed by a power-law decay, viz.~
\begin{align}
H(t)=\begin{cases}
\widetilde{H}&:t<\tbr\ ,\\
\widetilde{H}\left(\frac{t}{\tbr}\right)^{-\alpha}&:\tbr<t\ .
\end{cases}
    \label{eq:H(t)}
\end{align}
The normalization $\widetilde{H}$, break-time $\tbr$ and power-law exponent $\alpha>0$ depend on the energy source in question (see Fig.~\ref{fig:heating}).  In Appendix \ref{sec:powerlaw} we rederive several of the analytic expressions from the constant-heating case (Sec.~\ref{sec:constant}), for this more complex heating evolution.

\begin{figure}
\begin{center}
\includegraphics[width=85mm, angle=0, bb=0 0 277 211]{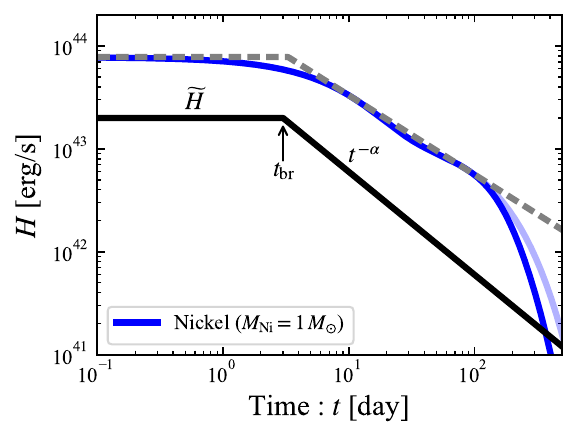}
\caption{The physically-motivated ejecta heating sources presented in this paper can in most cases be approximated as a constant heating rate $\widetilde{H}$ followed after a break at time $t_{\rm br}$ by a power-law decay $\propto t^{-\alpha}$ of the form (Eq.~\eqref{eq:H(t)}), shown here schematically with a black solid line. A gray dashed line denotes such a form fit to the $^{56}$Ni$\rightarrow ^{56}$Co$\rightarrow ^{56}$Fe decay chain, the latter shown as dark(light) solid blue lines in a case with(without) the gamma-ray energy deposition factor $f_{\rm dep}$.}
\label{fig:heating}
\end{center}
\end{figure}

\subsection{Radioactive decay heating}
It is well known that heating from radioactive decay extends the plateau duration in SNIIP \citep[e.g.,][]{Young2004,Kasen&Woosley2009,Bersten+2011,Nakar+2016,Goldberg+2019,Kozyreva+2019}. In particular, the heating rate of the $^{56}$Ni$\rightarrow ^{56}$Co$\rightarrow ^{56}$Fe decay chain can be written \citep{Nadyozhin1994}
\begin{align}
H(t)&=M_{\rm Ni}\left[\left(\varepsilon_{\rm Ni}-\varepsilon_{\rm Co}\right)e^{-t/t_{\rm Ni}}+\varepsilon_{\rm Co}e^{-t/t_{\rm Co}}\right]\ ,
    \label{eq:H(t)_Ni}
\end{align}
where $M_{\rm Ni}$ is the nickel mass, $\varepsilon_{\rm Ni}=3.9\times10^{10}\,\rm erg\,g^{-1}\,s^{-1}$, $\varepsilon_{\rm Co}=6.8\times10^{9}\,\rm erg\,g^{-1}\,s^{-1}$, $t_{\rm Ni}=8.8\,\rm day$, and $t_{\rm Co}=111.3\,\rm day$. The total energy released by radioactive decay,
\begin{align}
E_{\rm Ni}&\equiv\int_{0}^\infty H(t)dt\simeq M_{\rm Ni}\left[t_{\rm Ni}\left(\varepsilon_{\rm Ni}-\varepsilon_{\rm Co}\right)+t_{\rm Co}\varepsilon_{\rm Co}\right]\ ,
	\nonumber\\
&\simeq1.8\times10^{48}{\,\rm erg\,}\left(\frac{M_{\rm Ni}}{10^{-2}\,\Msun}\right)\ ,
\end{align}
is typically much smaller than the ejecta kinetic energy for values $M_{\rm Ni} \sim 10^{-3}-0.1M_{\odot}$ characteristic of SNIIP \citep[e.g.,][]{Hamuy2003,Anderson+2014,Valenti+2016,Muller+2017,Anderson2019,Martinez+2022b}. Because ejecta acceleration by radioactive decay energy is generally negligible, it is safe to assume constant ejecta speed in analytic estimates.

Most of the energy released by the $^{56}$Ni decay chain is carried by gamma-rays, which at sufficiently late times leak out of the ejecta without depositing energy. This suppression effect on the heating rate (Eq.~\eqref{eq:H(t)_Ni}) can be included by a multiplying deposition factor \citep{Jeffery1999}:
\begin{align}
f_{\rm dep}(t)&=1-e^{-x\tau_{\gamma}}\ ,
\label{eq:fdep(t)_Ni}
\end{align}
where $\tau_{\gamma}=\kappa_{\gamma}\rho R$ is the gamma-ray optical depth and $\kappa_{\gamma}=0.03\,\rm cm^{2}\,g^{-1}$ the associated opacity \citep{Swartz+1995}. After the ejecta becomes optically thin to gamma-rays at $\simeq(\kappa_{\gamma}/\kappa)^{1/2}t_{\rm thin}\simeq230\,\rm days$, the deposition efficiency steeply declines $f_{\rm dep}(t)\propto t^{-2}$.

\begin{figure}
\begin{center}
\includegraphics[width=85mm, angle=0, bb=0 0 278 398]{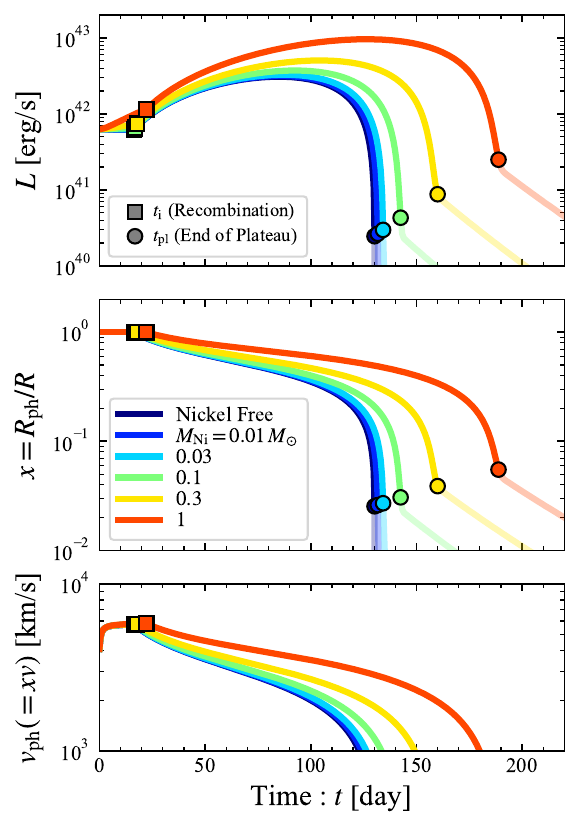}
\caption{Time evolution of the bolometric luminosity, location and velocity of the photosphere (similar to Fig.~\ref{fig:evolution_h}) for models incorporating the exact $^{56}$Ni decay chain heating, for different values of the $^{56}$Ni mass as marked.}
\label{fig:evolution_ni}
\includegraphics[width=85mm, angle=0, bb=0 0 282 227]{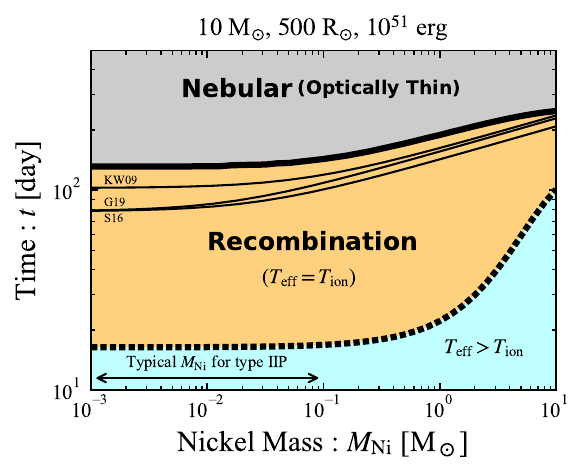}
\caption{The same as Fig.~\ref{fig:time_h}, but for $^{56}$Ni decay chain heating. Thin solid lines show analytic estimates for the plateau duration from the literature \citep[from top to bottom]{Kasen&Woosley2009,Goldberg+2019,Sukhbold+2016}. A double-headed arrow denotes the typical range of $^{56}$Ni masses inferred for SNIIP, $M_{\rm Ni}\simeq10^{-3}-0.1\,\Msun$.}
\label{fig:time_ni}
\end{center}
\end{figure}

Fig.~\ref{fig:evolution_ni} depicts the time evolution of the same quantities (luminosity, photosphere radius, photosphere velocity) shown in Fig.~\ref{fig:evolution_h} for the time-dependent $H(t)$ and $f_{\rm dep}(t)$ described by Eqs.~\eqref{eq:H(t)_Ni} and \eqref{eq:fdep(t)_Ni}. As expected, the recombination phase grows longer for larger $^{56}$Ni masses. However, different from the case of constant heating, the ejecta becomes optically thin prior to when the luminosity begins to track the input heating rate (i.e., the timescale $t_{\rm diff}^{\rm rec}$ is not reached) because of the sharp drop-off in the heating rate at $t \gtrsim100\,\rm days$ (Fig.~\ref{fig:heating}).

Fig.~\ref{fig:time_ni} shows the key timescales from Fig.~\ref{fig:time_h} but now as a function of $^{56}$Ni mass instead of generic heating rate. Although we consider nickel masses which range up to the total ejecta mass ($M=10\,\Msun$), such extreme yields of radioactive material are obviously unphysical, except perhaps in some PISNe.  For nickel masses typical of SNIIP, the onset time of the recombination phase $\ti$ is essentially unaffected by radioactive heating. Though still relatively modest, lengthening of the plateau duration with increasing $^{56}$Ni mass, is more readily apparent. 

The above results obtained employing the exact heating rate (Eq.~\eqref{eq:H(t)_Ni}) can also be understood quantitatively within our analytical framework, following Appendix \ref{sec:powerlaw}. Fig.~\ref{fig:heating} shows that the heating rate of the $^{56}$Ni decay chain at $t \lesssim100\,\rm days$ is well-approximated as a broken power-law function of the form Eq.~\eqref{eq:H(t)}, with:
\begin{align}
\widetilde{H}&=M_{\rm Ni}\varepsilon_{\rm Ni}\simeq7.8\times10^{41}{\,\rm erg\,s^{-1}\,}\left(\frac{M_{\rm Ni}}{10^{-2}\,\Msun}\right)\ ,
    \nonumber\\
\tbr&=3.3{\,\rm day}\ ,
    \nonumber\\
\alpha&=0.77\ .
    \nonumber
\end{align}
Neglecting the gamma-ray deposition factor, our analytic estimate for the plateau duration for these parameters follows from Eq.~\eqref{eq:tpl4},
\begin{align}
\tpl\simeq200{\,\rm day\,}\left(\frac{M_{\rm Ni}}{0.1\,\Msun}\right)^{0.15}E_{0,51}^{-0.44}M_{10}^{0.74}\ ,
    \label{eq:tpl_Ni}
\end{align}
where we have omitted the dependence on $T_{\rm i}$, and eliminated the $v-$dependence using Eq.~\eqref{eq:Ekin} and by equating the initial internal and kinetic energies $E_0=E_{\rm kin,0}$. By equating Eq.~\eqref{eq:tpl_Ni} with Eq.~\eqref{eq:tpl_popov}, we obtain the minimal nickel mass to modify the plateau duration:
\begin{align}
M_{\rm Ni,min}\simeq8.1\times10^{-3}\,\Msun\,R_{0,500}^{0.97}E_{0,51}^{1.8}M_{10}^{-1.4}\ .
    \label{eq:Mni_min}
\end{align}

We now compare these results with previous works. \cite{Kasen&Woosley2009} present a heuristic argument for how to include $^{56}$Ni heating as a correction to the analytic plateau duration of \citet{Popov1993}, described by: 
\begin{align}
t_{\rm pl}=t_{\rm pl,Popov}\times\left[1+C_{\rm f}\left(\frac{M_{\rm Ni}}{\Msun}\right)E_{\rm SN,51}^{-1/2}M_{10}^{-1/2}R_{0,500}^{-1}\right]^{1/6}\ ,
    \label{eq:tpl_KW09}
\end{align}
where $E_{\rm SN}(=E_0+E_{\rm kin,0})$ is the total explosion energy. The constant $C_{\rm f}$ is sensitive to the ejecta density profile and distribution of $^{56}$Ni in the ejecta, and can be calibrated by radiation transport simulations, with \citet{Kasen&Woosley2009} finding $C_{\rm f}\simeq21$, \citet{Sukhbold+2016} finding $\simeq 50$, and \cite{Goldberg+2019} finding $\simeq87$. Eq.~\eqref{eq:tpl_KW09} is shown as thin solid curves in Fig.~\ref{fig:time_ni} for different values of $C_{\rm f}$. Here, the nickel-free plateau duration $t_{\rm pl,Popov}$ is not given by Eq.~\eqref{eq:tpl_popov} but rather is taken directly from each reference (see the footnote at the end of Sec.~\ref{sec:method}). Our result (thick solid curve) is broadly consistent with these numerically-calibrated analytic expressions. 

While our results and previous studies show reasonable agreement, there are differences between them. From Eq.~\eqref{eq:tpl_KW09}, the minimum nickel mass required to impact the plateau duration can estimated by equating the second term in the square bracket with unity:
\begin{align}
M_{\rm Ni,min}\simeq4.8\times10^{-2}\,\Msun\,\left(\frac{C_{\rm f}}{21}\right)^{-1}E_{\rm SN,51}^{1/2}M_{10}^{1/2}R_{500}\ .
    \label{eq:Mni_min_KW09}
\end{align}
For nickel masses larger than this value (which corresponds to a $2^{1/6}\simeq$12\% increase in $t_{\rm pl}$ versus the Ni-free case), the plateau duration is considerably increased, now scaling with the ejecta properties according to: 
\begin{align}
t_{\rm pl}\propto M_{\rm Ni}^{1/6}E_{\rm SN}^{-1/4}M^{5/12}\simeq M_{\rm Ni}^{0.17}E_{\rm SN}^{-0.25}M^{0.42}\ ,
    \label{eq:tpl_KW09_2}
\end{align}
where we used Eq.~\eqref{eq:tpl_ref} for $t_{\rm pl,Popov}$. The exponents on the ejecta mass and explosion energy here differ from our analytical expressions in Eqs.~\eqref{eq:tpl_Ni} and \eqref{eq:Mni_min}, with in particular the dependence of $M_{\rm Ni,min}$ on $M$ changing sign. This disagreement results from the different ways that $^{56}\rm{Ni}$ heating is treated in our formula versus that of \citet{Kasen&Woosley2009}. We define $t_{\rm pl}$ by $x\tau=1$, and account for $^{56}\rm{Ni}$ heating during the plateau phase, which delays the recession of the recombination front $x(t)$. By contrast, \citet{Kasen&Woosley2009} consider the heat from radioactive decay to boost the internal energy such that heating effectively occurs immediately at $t_{\rm Ni}$ and $t_{\rm Co}$ (see their Eq.~(9)). Therefore in our formulation increasing $M$ slows the intrinsic evolution of $x$, thus allowing a lower $M_{\rm Ni}$ to significantly increase $t_{\rm pl} > t_{\rm pl,Popov}$. In the \citet{Kasen&Woosley2009} scenario, increasing $M$ boosts the internal energy available at a given $t$ ($> t_{\rm Ni}$ or $t_{\rm Co}$) due to the smaller adiabatic losses experienced for lower expansion velocity (at fixed $E_{\rm SN}$), which thus requires higher $M_{\rm Ni}$ to generate enough heating to boost $t_{\rm pl} > t_{\rm pl,Popov}$. We remark that the exponent of $M$ in our $M_{\rm Ni,min}$ depends on $\alpha$ as $M_{\rm Ni,min}\propto M^{\frac{5(3\alpha-10)}{28}}$.

\subsection{Magnetar}\label{sec:magnetar}

Another frequently invoked central energy source is the spin-down energy of a newly-formed, rapidly spinning magnetized neutron star (``magnetar''; e.g., \citealt{Maeda+2007,Kasen&Bildsten2010,Woosley2010,Chatzopoulos+2012,Metzger+2015d,Sukhbold&Thompson2017}).  Assuming the Poynting luminosity of the magnetar wind, and the high-energy radiation it generates, to be thermalized within the ejecta (likely a good approximation at least at early times; \citealt{Vurm&Metzger2021}), the heating rate evolution will follow the standard magnetic dipole spin-down rate (e.g., \citealt{Spitkovsky2006}). The latter can be expressed in the form:
\begin{align}
H(t)&=\frac{E_{\rm rot}}{t_{\rm sd}}\left(1+\frac{t}{t_{\rm sd}}\right)^{-2}\ ,
    \label{eq:H(t)_NS}
\end{align}
where $E_{\rm rot}$ and $t_{\rm sd}$ are the magnetar's initial rotational energy and spin-down timescale, respectively. These quantities are roughly related to the initial spin period $P_{\rm i}$ and surface magnetic field strength $B$ by 
\begin{align}
P_{\rm i}&\simeq\left(\frac{2\pi^2I}{E_{\rm rot}}\right)^{1/2}\simeq1.4{\,\rm ms\,}E_{\rm rot,52}^{-1/2}\ ,\\
B&\simeq\left(\frac{c^3I^2}{4 R_{\rm NS}^6E_{\rm rot}t_{\rm sd}}\right)^{1/2}\simeq8.8\times10^{13}{\,\rm G\,}E_{\rm rot,52}^{-1/2}\left(\frac{t_{\rm sd}}{\rm day}\right)^{-1/2}\ ,
\end{align}
where $I$ and $R_{\rm NS}$ are the magnetar's moment of inertia and radius, respectively, and we adopt $I=10^{45}\,\rm g\,cm^2$ and $R_{\rm NS}=10\,\rm km$.
However, we neglect this correspondence for simplicity, instead referring directly to $t_{\rm sd}$ and $E_{\rm rot}$. One hard constraint is that a magnetar cannot spin faster than its centrifugal break-up rate, which limits $E_{\rm rot} \lesssim 10^{53}\,\rm erg$ \citep{Metzger+2015d}.
Additionally, different from heating due to radioactive decay or CSM interaction, energy input from the magnetar can appreciably accelerate the supernova ejecta (e.g., \citealt{Kasen+2016,Suzuki&Maeda2021}).

\begin{figure}
\begin{center}
\includegraphics[width=85mm, angle=0, bb=0 0 285 225]{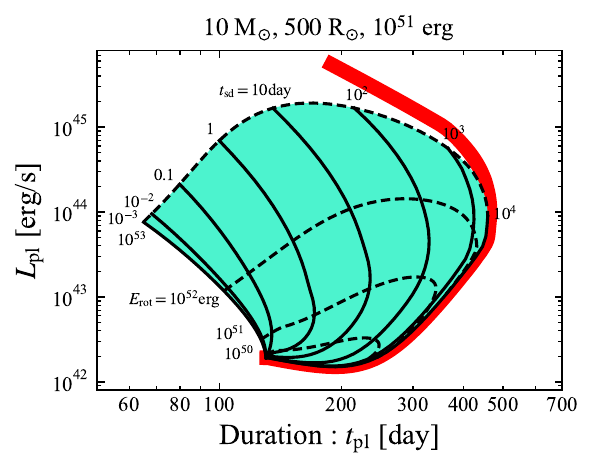}
\caption{The same as Fig.~\ref{fig:tL_h} but for heating by magnetar spin-down. Black solid (dashed) curves show contours for fixed spin down timescale, $t_{\rm sd}$ (and fixed magnetar's rotation energy, $E_{\rm rot}$). A red thick curve shows the constant heating case (see Fig.~\ref{fig:tL_h}), which bounds the high-$t_{\rm pl}$ edge of the parameter space.}
\label{fig:tL_mag}
\end{center}
\end{figure}

Fig.~\ref{fig:tL_mag} shows the plateau duration and luminosity for a range of values $E_{\rm rot}=10^{48}-10^{53}\,\rm erg$ and $t_{\rm sd}=10^{-3}-10^{4}\,\rm day$, corresponding to physically allowed birth spin periods and magnetic field strengths. At a fixed spin down time, the plateau luminosity increases monotonically with $E_{\rm rot}$, while the plateau duration initially grows but then turns over and decreases for $E_{\rm rot}\gtrsim E_{\rm kin,0}=10^{51}\,\rm erg$ once acceleration of the ejecta becomes significant.

Contours of fixed $t_{\rm sd}$ accumulate in the two extreme limits $t_{\rm sd}\to 0$ and $\infty$. 
For spin down times much shorter than the initial dynamical time $t_{\rm sd}\lesssim t_0\sim1\,\rm day$, heating of the ejecta is effectively instantaneous. This leads to acceleration of the ejecta at the expense of the deposited energy for $t\gtrsim t_0$, with the subsequent evolution practically described by replacing $E_0\to E_{\rm rot}$ and $v\to\sqrt{10E_{\rm rot}/3M}$ in the Popov formula. Because for early energy injection the ejecta loses memory of when energy is deposited, the contours becomes independent of $t_{\rm sd}$ for small $t_{\rm sd}\lesssim10^{-2}\,\rm day$. 

In the opposite limit $t_{\rm sd}\to \infty$, the contours asymptote to the constant heating case. For heating times longer than the plateau duration $t_{\rm sd}\gtrsim t_{\rm pl,Popov}\sim10^{2}\,\rm day$, the magnetar provides what is effectively a constant heating source during the plateau. A red curve shows the constant heating rate limit (see Fig.~\ref{fig:tL_h}), which bounds the allowed plateau duration in the magnetar model. 

Most previous applications of magnetar engines are to hydrogen-poor superluminous supernovae; however, a few works have explored the impact of a magnetar on hydrogen-rich supernova light curves using numerical calculations \citep{Sukhbold&Thompson2017,Dessart&Audit2018,Orellana+2018}. While these works adopt different definitions for the plateau duration and luminosity, precluding a quantitative comparison, their findings are broadly compatible with the simpler analytic estimates presented here.

\subsection{Accreting Compact Object}
Another potential heating source of the supernova ejecta is a relativistic jet\footnote{For red supergiants, their massive thick envelopes prevent a jet from breaking out of the stellar surface and producing high-energy emission akin to typical long gamma-ray bursts (e.g., \citealt{Matzner2003,Matsumoto+2015} but see \citealt{Quataert&Kasen2012} for the case of very-long duration events).} or accretion disk outflow, powered by mass accretion on the central black hole or neutron star \citep[e.g.,][]{Dexter&Kasen2013,Moriya+2018e,Kaplan&Soker2020}. The rate of fall-back accretion is expected to decline at late times as $\dot{M}_{\rm fb}\propto t^{-5/3}$ \citep{Michel1988,Chevalier1989}, following an initial phase defined by the radial structure of the progenitor star \citep{Dexter&Kasen2013,Perna+14,Moriya+2019}, particularly its radius (e.g., \citealt{Perna+18}).  The manner in which the mass fall-back is processed by the central accretion disk or its outflows, can also modify the accretion rate reaching the central compact object and hence the heating rate \citep[e.g.,][]{Metzger+2008c}. We encapsulate these uncertainties by adopting a broken power-law heating rate, similar to the magnetar case:
\begin{align}
H(t)=\frac{(\alpha_{\rm fb}-1)E_{\rm acc}}{t_{\rm fb}}\left(1+\frac{t}{t_{\rm fb}}\right)^{-\alpha_{\rm fb}}\ ,
	\label{eq:H(t)_BH}
\end{align}
where the total available energy $E_{\rm acc}$ now scales with the accreted mass $M_{\rm acc}$ for some assumed efficiency $\eta = E_{\rm acc}/M_{\rm acc}c^2$, and $t_{\rm fb}$ is the characteristic fallback time.  The latter is generally expected to scale with the fall-back time of the stellar envelope, 
\begin{align}
t_{\rm fb}\sim \frac{1}{\sqrt{G\rho_\star}}\simeq130{\,\rm day\,}\left(\frac{M_\star}{10\,\Msun}\right)^{-1/2}\left(\frac{R_\star}{500\,\Rsun}\right)^{3/2}\ ,
\end{align}
where $\rho_{\star} \equiv 3M_{\star}/4\pi R_{\star}^{3}$ is the mean density of the star of mass $M_\star$ and radius $R_{\star}$. We consider the post-break exponent $\alpha_{\rm fb}$, as a free parameter. The $\alpha_{\rm fb}-1$ prefactor in Eq.~\eqref{eq:H(t)_BH} follows from the normalization $E_{\rm acc}=\int_0^\infty H(t)dt$.

\begin{figure}
\begin{center}
\includegraphics[width=85mm, angle=0, bb=0 0 285 225]{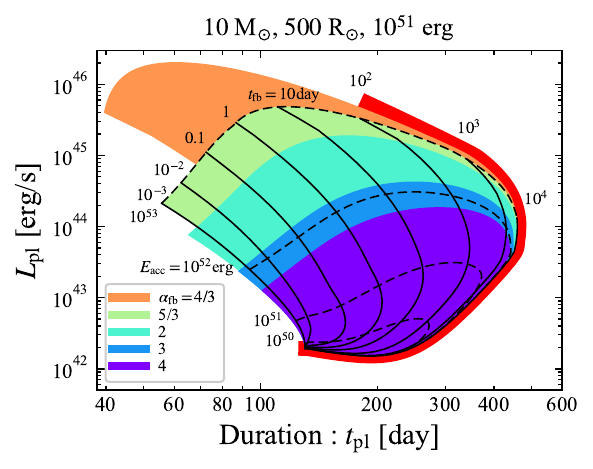}
\caption{The same as Fig.~\ref{fig:tL_h} but for heating by accretion onto a central compact object, following Eq.~\eqref{eq:H(t)_BH}. Each shaded region corresponds to the parameter space of plateau duration and luminosity for different values of the post-break index $\alpha_{\rm fb}$. Shaded regions are overlaid such that the curves for larger $\alpha_{\rm fb}$ cover part of the smaller $\alpha_{\rm fb}$ curves. All values of $\alpha_{\rm fb}$ are bounded at the bottom by the constant-heating case. The region for $\alpha_{\rm fb}=2$ essentially follows the magnetar heating case (see Fig.~\ref{fig:tL_mag}). Black solid and dashed contours denote cases of fixed fallback timescale and total accretion energy for $\alpha_{\rm fb}=5/3$. A red thick curve shows the case of a constant heating rate (see Fig.~\ref{fig:tL_h}), which bounds the high-$t_{\rm pl}$ edge of the parameter space (see also Fig.~\ref{fig:tL_h}).}
\label{fig:tL_bh}
\end{center}
\end{figure}

Fig.~\ref{fig:tL_bh} depicts the plateau duration and luminosity for different $\alpha_{\rm fb}$, $E_{\rm acc}$, and $t_{\rm fb}$, similar to the format of Fig.~\ref{fig:tL_mag}. The colors of each shaded region correspond to the range of $t_{\rm pl}$ and $L_{\rm pl}$ for different ranges of $\alpha_{\rm fb}=4/3$ to $4$ as marked. The contours follow a similar shape to that of the magnetar heating (corresponding to $\alpha_{\rm fb}=2$). Black solid and dashed curves represent the contours for fixed $E_{\rm acc}$ and $t_{\rm fb}$ for the special case $\alpha_{\rm fb}=5/3$ (corresponding to a canonical fallback rate). For smaller exponents, the heating rate declines more slowly, increasing the plateau luminosity. As in the case of magnetar heating, the plateau duration is maximized for longer fallback timescales, but the attainable region is bounded by the limit of a constant heating rate. 

\subsection{CSM Shock Interaction}
\label{sec:shock interaction}

\begin{figure}
\begin{center}
\includegraphics[width=85mm, angle=0,bb=0 0 1715 850]{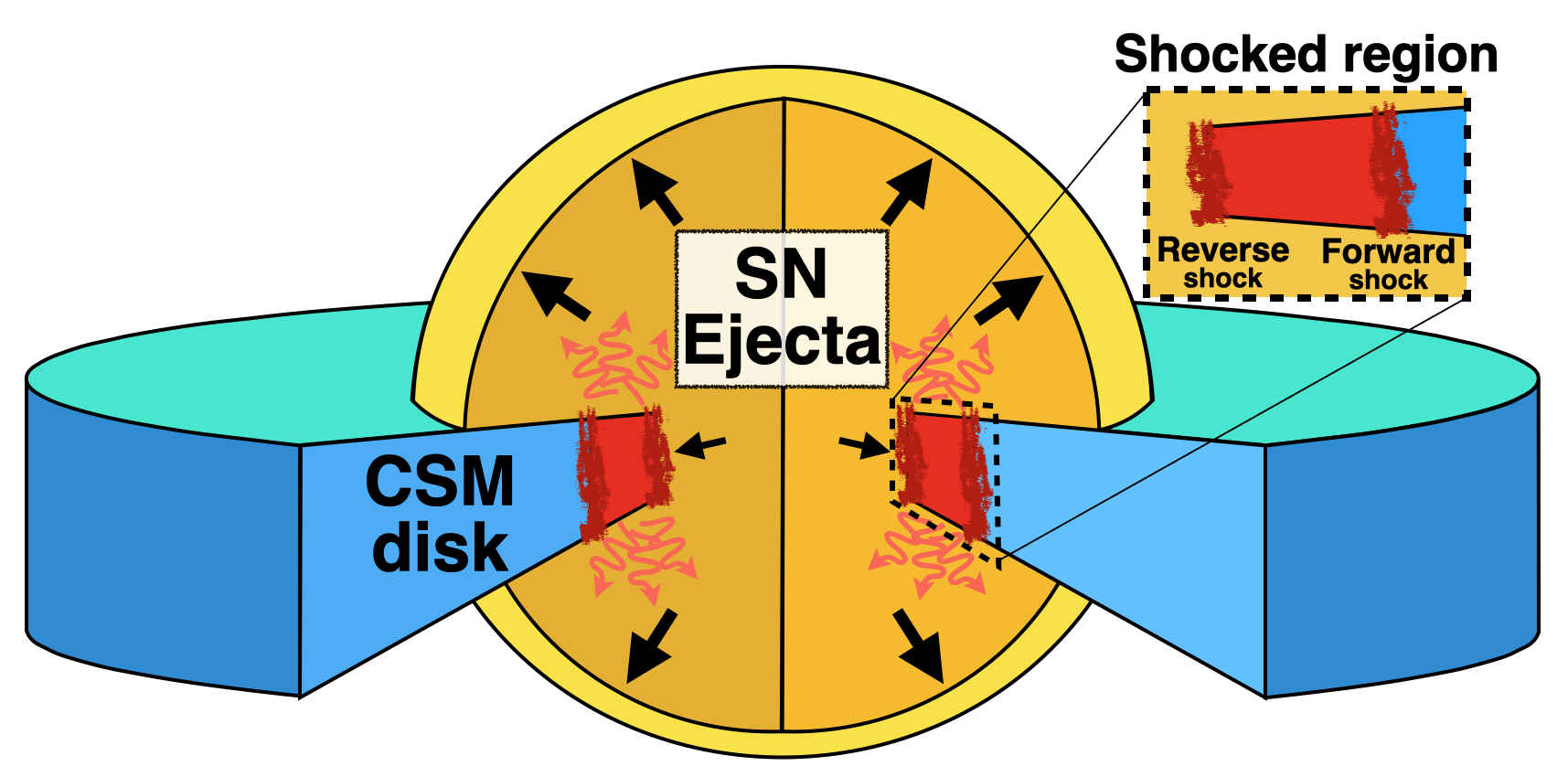}
\caption{A schematic diagram of the envisioned scenario for shock interaction between the supernova ejecta and an equatorial CSM disk.  Radiation released by the ejecta-CSM shocks in the equatorial regions provide a source of internal (sub-photospheric) heating for bulk of the faster, effectively freely-expanding ejecta at higher latitudes. }
\label{fig:picture}
\end{center}
\end{figure}

As a final heating source, we consider shock interaction between the supernova ejecta and a circum-stellar medium (CSM). While most of the modeling literature focuses on CSM which distributed spherically symmetrically around the explosion site \citep[e.g.,][]{Khatami&Kasen2024}, this geometry confines effects of shock heating to the outermost layers of the ejecta. While shock heating can in that case contribute to the early-time supernova light curve by impacting the initial thermal energy imparted to these outer layers, it does not provide a sustained heating source embedded behind the photosphere which would affect the duration of the plateau and later-time SN evolution. We are thus motivated to consider the case of aspherical CSM, confined into a disk or otherwise equatorially-focused configuration surrounding the progenitor \citep{Blondin+1996,vanMarle+2010,Vlasis+2016,McDowell+2018,Kurfurst&Krticka2019,Suzuki+2019b,Kurfurst+2020,Nagao+2020}, a geometry which is supported by some observations of type IIn or otherwise peculiar SNe  \citep{Chugai&Danziger1994,Leonard+00,Andrews+17,Andrews&Smith2018,Andrews+19,Bilinski+20,Bilinski+2024}. Indeed, the most luminous SLSNe-II show no early-time evidence for narrow lines (e.g., \citealt{Gezari+09,Miller+09}), supporting a configuration in which the CSM shock is initially embedded behind the photosphere. Additionally, a fraction of SLSNe-II show no narrow lines during photospheric phase \citep{Inserra+2018c,Kangas+2022}.

Fig.~\ref{fig:picture} provides a schematic depiction of the envisioned shock interaction. The supernova ejecta collides with a CSM disk starting at its inner edge, forming forward and reverse shocks, sandwiched between a swept-up shell of shocked gas, which expands outwards in time.  Since the portion of the supernova ejecta directed along the polar region expands relatively freely without encountering significant mass, it can envelope the slower-expanding shocked region (e.g., \citealt{Metzger10,Andrews&Smith2018}). Radiation generated at the equatorial shock thus diffuses vertically into the ejecta, serving as an effective internal heating source for a (one-dimensional) supernova light curve model (e.g., \citealt{Metzger&Pejcha2017}). We assume that the shocked CSM expands only into the radial direction, retaining its initial solid angle.  
We also neglect the finite time required for radiation to diffuse radially out of the disk and into the surrounding supernova ejecta; our model thus provides a maximal heating luminosity and hence a conservative upper limit on the corresponding heating-extended plateau duration.

We adopt a radial density profile for the equatorial disk corresponding to that of a steady-wind,
\begin{align}
\rho_{\rm CSM}=\frac{\dot{M}}{4\pi f_{\Omega}v_{\rm CSM}r^2}\ ,
    \label{eq:CSM_profile}
\end{align}
where $f_\Omega < 1$ is the fraction of total solid angle subtended by the disk, $v_{\rm CSM}$ is the CSM radial velocity, and $\dot{M}$ is the wind mass-loss rate. We estimate the effective heating rate due to CSM shock interaction as follows (a more detailed description of the shock evolution, for a generic density profile index, is given in Appendix \ref{sec:shock2}). The time evolution of the shocked region can be described by mass and momentum conservation \cite[e.g.,][]{Metzger&Pejcha2017}.  We assume sufficiently dense CSM that both forward and reverse shocks are radiative.  Since the radial width of the shocked region is small in this case, we characterize the shell of shocked gas by a representative radius $R_{\rm sh}$ and velocity $v_{\rm sh}$.

Initially, the shock expands outwards at a roughly constant speed close to that of the supernova ejecta, the swept-up mass growing as $M_{\rm sh} \simeq \dot{M}vt/v_{\rm CSM}.$  Once $M_{\rm sh}$ becomes comparable to the shocked ejecta mass, $f_\Omega M$, the shell of shocked gas starts to appreciably decelerate, on a characteristic timescale
\begin{align}
t_{\rm dec}\simeq\frac{f_\Omega M v_{\rm CSM}}{\dot{M}v}&\simeq610{\,\rm day\,}\frac{f_{\Omega,-1} M_{10}v_{\rm CSM,100}}{v_{6000}\dot{M}_{-2}^{-1}}\ ,
\end{align}
where $v_{\rm CSM}=100\,v_{\rm CSM,100}\,\rm km\,s^{-1}$, $\dot{M}=10^{-2}\dot{M}_{-2}\,\Msun\,\rm yr^{-1}$, and $f_{\Omega}=0.1f_{\Omega,-1}$. From momentum conservation, the radius and velocity of the shell evolve according to,
\begin{align}
R_{\rm sh}&\simeq\begin{cases}
vt&: t<t_{\rm dec}\ ,\\
vt_{\rm dec}\left(\frac{t}{t_{\rm dec}}\right)^{1/2}&: t>t_{\rm dec}\ ,
\end{cases}\\
v_{\rm sh}&\simeq\begin{cases}
v&: t<t_{\rm dec}\ ,\\
v\left(\frac{t}{t_{\rm dec}}\right)^{-1/2}&: t>t_{\rm dec}\ .
\end{cases}
\end{align}
The kinetic luminosity of the forward shock dominates that of the reverse shock, and thus the forward shock dominates the ejecta heating. The forward shock luminosity evolves according to:
\begin{align}
H(t)&\sim4\pi f_{\Omega}R_{\rm sh}^2\rho_{\rm CSM}v_{\rm sh}^3=\frac{\dot{M}v_{\rm sh}^3}{v_{\rm CSM}}
    \label{eq:H(t)_shock}\\
&\simeq1.4\times10^{43}{\,\rm erg\,s^{-1}\,}\frac{v_{6000}^3\dot{M}_{-2}}{v_{\rm CSM,100}}\begin{cases}
1&: t<t_{\rm dec}\ ,\\
\left(\frac{t}{t_{\rm dec}}\right)^{-\frac{3}{2}}&: t>t_{\rm dec}\ .
\end{cases}
	\nonumber
\end{align}
Interestingly then, for a wind-like CSM profile $\rho_{\rm CSM} \propto r^{-2}$, the heating rate evolution again follows the functional form of Eq.~\eqref{eq:H(t)}.

Our treatment of the shock as an internal heating source requires the shocked region to remain inside the supernova photosphere, i.e.,
\begin{align}
R_{\rm sh}\leq R_{\rm ph}(=xR)\ .
    \label{eq:Rsh<Rph}
\end{align}
After this condition becomes violated, we truncate the heating rate in our model, i.e. $H = 0$ for $R_{\rm sh}>R_{\rm ph}$.

\begin{figure}
\begin{center}
\includegraphics[width=85mm, angle=0, bb=0 0 286 398]{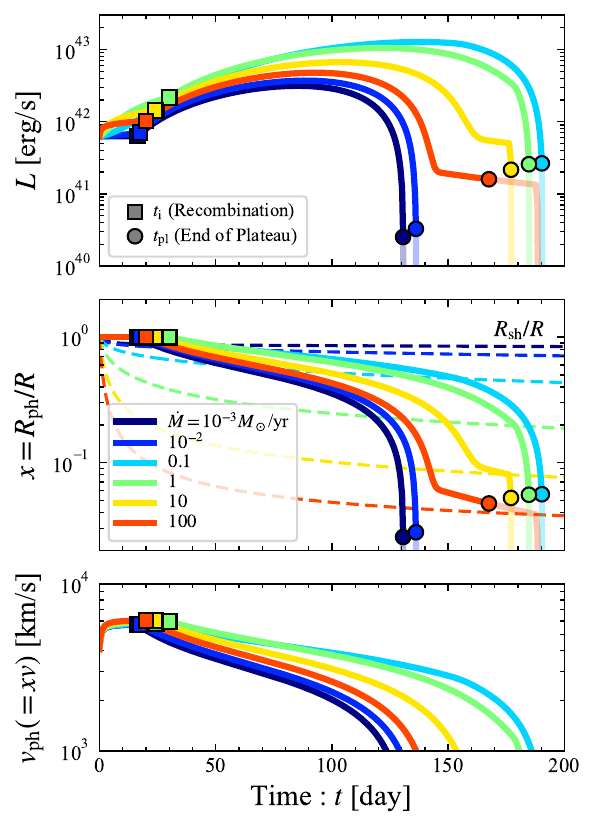}
\caption{The same as Fig.~\ref{fig:evolution_ni} but for energy injection due to shock interaction between the supernova ejecta and a disk-like CSM. For the CSM velocity and solid angle covering fraction, we adopt $v_{\rm CSM}=100\,\rm km\,s^{-1}$ and $f_{\Omega}=0.1$, respectively. In the middle panel, dashed curves show the shock radius in units of the ejecta radius, $R_{\rm sh}/R$. Once the shocks emerge from the photosphere such that $R_{\rm sh} > R_{\rm ph}$ (intersection of solid and dashed curves), the shock can no longer supply a source of internal heating to the ejecta and internal heating is truncated. This truncation can best be seen in the yellow line ($\dot{M}=10M_\odot/$yr).}
\label{fig:evolution_shock}
\end{center}
\end{figure}

Fig.~\ref{fig:evolution_shock} shows light curves, as well as the evolution of photosphere radius and velocity, for various assumed CSM wind mass-loss rates, obtained by calculating the shock evolution numerically (see Appendix \ref{sec:shock2} for the exact equations). The inner edge of the CSM disk is assumed to coincide with the progenitor star radius, and we fix the other CSM properties according to $f_{\Omega}=0.1$ and $v_{\rm CSM}=100\,\rm km\,s^{-1}$ (the results depend only on the overall CSM density normalization $\propto \dot{M}/v_{\rm CSM}$, provided that $v_{\rm CSM}\ll v_{\rm sh}$). As a general rule, the ejecta cannot be appreciably accelerated as a result of CSM interaction because the ultimate source of the ejecta heating is the initial kinetic energy of the ejecta which intercepts the CSM, $\sim f_\Omega E_{\rm kin,0}$. 

While the lengthening of the plateau phase from shock heating is typically only moderate, we identify an ``optimal'' pre-explosion progenitor mass-loss rate, $\dot{M}\sim0.1\,\Msun\,\rm yr^{-1}$ (for $v_{\rm CSM} = 100$ km s$^{-1}$), which maximizes the plateau duration. While the shock power increases for larger $\dot{M}$, the duration of this heating phase is correspondingly shorter because of the shorter deceleration time, which results in a lower shock luminosity during the recombination phase. On the other hand, for less massive CSM (smaller $\dot{M}$), the shock heating is sustained for a longer timescale, but at a value too low to sustain the ejecta ionization and appreciably lengthen the plateau. The shocked region also emerges faster from the photosphere in the freely-coasting case, violating condition \eqref{eq:Rsh<Rph} and terminating the ejecta heating.

\begin{figure}
\begin{center}
\includegraphics[width=85mm, angle=0,bb=0 0 282 231]{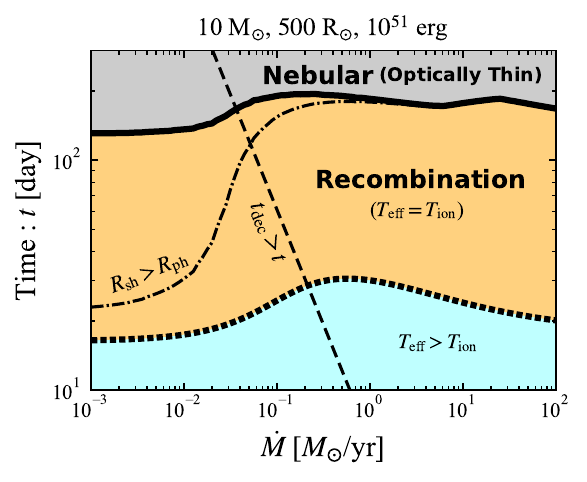}
\caption{The same as Fig.~\ref{fig:time_h} but for shock interaction heating for different normalizations of the wind-like CSM density profile. Dashed diagonal lines denote the deceleration timescale before which the shock expands freely and the shock luminosity is approximately constant. A dash-dotted line shows the time at which the shock radius becomes larger than the retreating supernova photosphere.}
\label{fig:time_shock}
\end{center}
\end{figure}

The above description of the optimal mass loss rate is supported by Fig.~\ref{fig:time_shock}, which again shows the key timescales, this time as a function of $\dot{M}$. For $\dot{M}\gtrsim 0.1\,\Msun\,\rm yr^{-1}$ deceleration of the ejecta is significant during the recombination phase (right of the dashed line), and resulting weaker shock power, $L_{\rm sh}\propto \dot{M}^{-1/2}$, causes the plateau duration to shorten. For $\dot{M}\lesssim0.1\,\Msun\,\rm yr^{-1}$, the shock does not experience significant deceleration during the recombination phase, and eventually overtakes the photosphere. The small rise in the plateau duration around $\dot{M}\sim10-100\,\Msun\,\rm yr^{-1}$ is caused by the tail of the light curve tracing the shock luminosity, as shown in Fig.~\ref{fig:evolution_shock}. In summary, the optimal mass loss rate for lengthening the plateau duration is that for which ejecta deceleration and recombination occur simultaneously. However, even in this optimal case, the plateau duration is extended by only a factor $\sim$2 compared to the no-heating case. In Appendix~\ref{sec:shock2}
we show this conclusion also holds for generic (i.e., non wind-like) CSM density profiles.

It is not straightforward to estimate analytically the optimal $t_{\rm pl}-$maximizing mass-loss rate because the effects of shock heating on the recombination process are subtle for $\dot{M}\sim10^{-2}-10^{-1}\,\Msun\,\rm yr^{-1}$. As an alternative, we adopt the mass-loss rate above which the resulting shock heating appreciably slows the ejecta recombination (the analog of $H_{\rm cr}$ from the constant heating rate case). Equating Eqs.~\eqref{eq:Hcr} with (\ref{eq:H(t)_shock}, $t<t_{\rm dec}$), we find: 
\begin{align}
\dot{M}_{\rm cr}\simeq3.4\times10^{-2}{\,\Msun\,\rm yr^{-1}\,}M_{10}v_{6000}^{-2}v_{\rm CSM,100}\left(\frac{\zeta}{0.83}\right)^{-3}\ .
    \label{eq:Mdot_cr}
\end{align}
Here, we have written $v_{\rm sh}=\zeta v$ to account for the moderate deceleration of the ejecta even prior to the formal deceleration time, where $\zeta\simeq 0.83$ is motivated by our numerical results. This estimate slightly underestimates the true $\dot{M}$ which maximizes $t_{\rm pl}$ in our full numerical calculations (Fig.~\ref{fig:time_shock}).

\section{Discussion}
\label{sec:discussion}

\begin{figure*}
\begin{center}
\includegraphics[width=150mm, angle=0,bb=0 0 322 258]{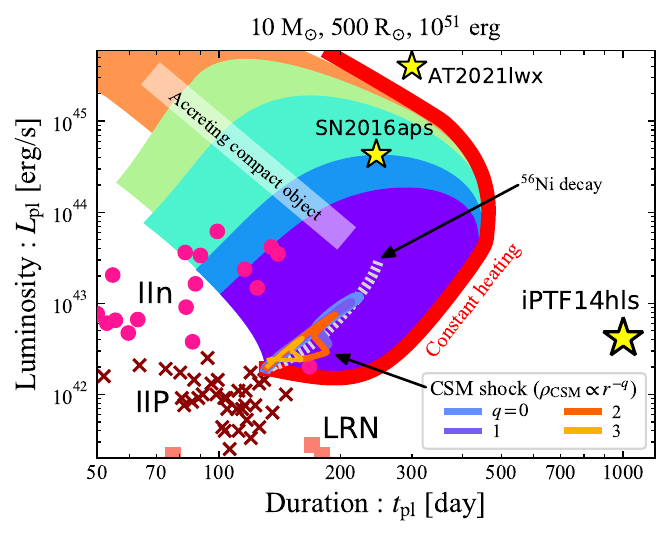}
\caption{Allowed space of plateau duration and luminosity for different heating sources. The colored shaded regions correspond to an accreting compact object (see Figs.~\ref{fig:tL_h}, \ref{fig:tL_bh}),
bounded by a solid red curve corresponding to constant heating-rate case.  Gray dashed and colored solid curves denote, respectively, radioactive $^{56}\rm Ni$ heating and CSM shock interaction (for a range of different radial density profiles). The ranges of nickel masses and CSM density normalizations follow those in Figs.~\ref{fig:time_ni} and \ref{fig:time_shock_v2}, respectively. Brown crosses, magenta circles, orange squares show the parameters for type IIP SNe, IIn SNe, and luminous red novae (LRNe) (data are taken from \citealt{Martinez+2022,Nyholm+2020,Matsumoto&Metzger2022b}). Three stars represent peculiar events, iPTF14hls \citep{Arcavi+2017d,Andrews&Smith2018}, SN 2016aps \citep{Nicholl+2020}, and AT 2021lwx \citep{Subrayan+2023,Wiseman+2023}. For type IIn, SN 2016aps, and AT 2021lwx, we use peak luminosities and show durations corresponding to the timescale the light-curve remains within $\lesssim 1$ magnitude of peak.}
\label{fig:tL}
\end{center}
\end{figure*}

Fig.~\ref{fig:tL} summarizes the allowed space of light curve properties, for the wide range of heating sources considered in this paper, for fiducial progenitor star/explosion properties. The presence of a sustained heating source generally acts to boost the plateau luminosity and duration above those of the no-heating case (the latter of which describes most SNIIP, shown for comparison as brown crosses). While increases in the luminosity of up to several orders of magnitude are possible for sufficiently powerful heating rates, the corresponding boost to the plateau duration is more tightly bounded, to within a factor $\lesssim 3$ of the duration in the zero-heating case.  The attainable plateau duration is capped by the limiting constant heating-rate case (depicted as a thick red curve). This maximal duration is approximately given by Eq.~\eqref{eq:tpl_max}, which expressed in terms of ejecta mass and explosion energy gives:
\begin{align}
t_{\rm pl,max} \simeq 580\,{\rm day}\,T_{\rm i,6000}^{-2/3}M_{10}^{17/24}E_{0,51}^{-7/24}\ .
\end{align}

Fig.~\ref{fig:tL} also shows for comparison a sample of type IIn supernovae, i.e. those showing narrow hydrogen lines indicative of CSM interaction, including the outlier event SN2016aps. We also show two other ``extreme'' explosions, iPTF14hls \citep{Arcavi+2017d} and AT2021lwx \citep{Subrayan+2023,Wiseman+2023}, though the latter has been interpreted as a tidal disruption event by some authors.

\begin{figure}
\begin{center}
\includegraphics[width=85mm, angle=0,bb=0 0 323 216]{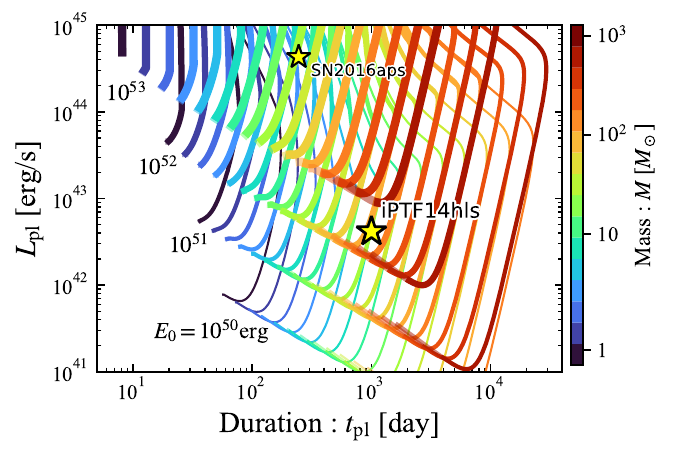}
\includegraphics[width=85mm, angle=0,bb=0 0 290 222]{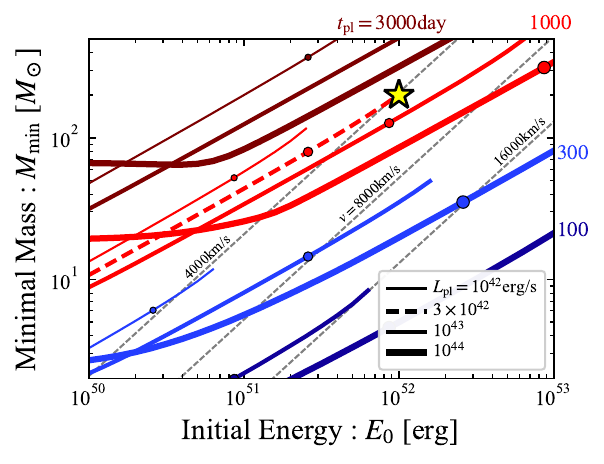}
\caption{({\bf Top}) Trajectories of plateau duration and luminosity for the bounding constant heating rate case ($H=const$), similar to Fig.~\ref{fig:tL_h} but now for a range of the energies $E_0=10^{50}\,\rm erg-10^{53}\,\rm erg$ and ejecta masses $M = 1-10^{3}\,\Msun$ as labeled. The range of heating rate values $H$ along each trajectory, is set to cover $H_{\rm min}$ (Eq.~\ref{eq:Hmin}) so that the heating-free solutions are present for each parameter set. On each trajectory, a segment violating the one-zone condition (Eq.~\ref{eq:M_violation}) is shown by transparent line. ({\bf Bottom}) Minimal ejecta mass $M_{\rm min}$ to produce a given plateau duration (different colored lines) and luminosity (different linestyles) as a function of the initial explosion energies $E_0$. Gray dashed lines show the corresponding mean ejecta velocity. Along each curve, a circle denotes the minimum energy in shock interaction scenarios, $E_0\gtrsim t_{\rm pl}L_{\rm pl}/f_{\Omega}$. A star represents the location of iPTF14hls with $t_{\rm pl}\simeq10^3\,\rm day$, $L_{\rm pl}\simeq3\times10^{42}\,\rm erg\,s^{-1}$, and $v\simeq4000\,\rm km\,s^{-1}$.}
\label{fig:tL2}
\end{center}
\end{figure}

\subsection{Minimum ejecta mass of long-plateau transients}\label{sec:min_mass}

The existence of a maximal plateau duration implies that a {\it minimum} ejecta mass is required to realize the plateau duration of a given observed event. Since the duration-luminosity trajectory for the limiting constant heating-rate case depends on the ejecta properties (bottom panel of Fig.~\ref{fig:tL_h}) we show in the top panel of Fig.~\ref{fig:tL2} these curves for an expanded parameter range of initial explosion energies $10^{50}\,\rm erg$ to $10^{53}\,\rm erg$ (thin to thick curves), and ejecta masses from $1$ to $10^3\,\Msun$ in steps of 0.2 dex.\footnote{We remove models satisfying $E_0=10^{50}\,\rm erg$ and $M \gtrsim 400\,\Msun$ because gas pressure dominates over radiation pressure in the ejecta for these parameters, violating our assumption (Eq.~\eqref{eq:Prad>Pgas}).} As expected from Sec.~\ref{sec:constant}, the maximal plateau-duration increases with ejecta mass and decreases with higher explosion energy.

The bottom panel of Fig.~\ref{fig:tL2} depicts the minimum ejecta mass, $M_{\rm min}$, needed to produce a given plateau duration. Based on the $L_{\rm pl}-t_{\rm pl}$ trajectory (Eq.~\ref{eq:Lpl_tpl}, for the $H<H_{\rm pl,max}$ case), we estimate:
\begin{align}
M_{\rm min}\simeq180\,\Msun\,E_{0,52}^{3/5}\left(\frac{L_{\rm pl}}{3\times10^{42}{\,\rm erg\,s^{-1}}}\right)^{-1/5}\left(\frac{t_{\rm pl}}{10^3{\,\rm day}}\right)^{6/5}\ ,
    \label{eq:Mminimal}
\end{align}
in rough agreement with the full calculation. If the ejecta velocity can be measured observationally, the explosion energy and hence $M_{\rm min}$ can be more robustly constrained. Along each curve, a small circle denotes the minimum explosion energy in shock-powered scenarios, $E_0\gtrsim t_{\rm pl}L_{\rm pl}/f_{\Omega}$, which is consistent with generating the radiated energy.

As expected from the $L_{\rm pl}-t_{\rm pl}$ trajectories, curves of $M_{\rm min}$ terminate at high energies (e.g., at $E_0\simeq10^{52}\,\rm erg$ for $L_{\rm pl}=10^{43}\,\rm erg\,s^{-1}$ and $t_{\rm pl}=300\,\rm day$), and flatten for lower energy (e.g., $E_0\lesssim10^{51}\,\rm erg$ for $L_{\rm pl}=10^{43}\,\rm erg\,s^{-1}$ and $t_{\rm pl}=1000\,\rm day$). The former behavior occurs because the $L_{\rm pl}-t_{\rm pl}$ trajectory rises with increasing $E_0$, until it no longer touches the given $L_{\rm pl}$. The flattening at low $E_0$ occurs because at high heating rates $H\gtrsim H_{\rm pl,max}$, the $L_{\rm pl}-t_{\rm pl}$ trajectory loses sensitivity to the initial explosion energy (Eq.~\ref{eq:Lpl_tpl} for the $H > H_{\rm pl,max}$ cases). This flattening occurs once the ejecta experiences significant acceleration, leading to a breakdown of the relations $E_0=3Mv^2/20$, $E_{\rm kin,0}=E_0$ and $v=\sqrt{2}v_0$ (Eqs.~\eqref{eq:Ekin}). Contours of constant ejecta velocity are denoted by gray dashed diagonal lines in the bottom panel of Fig.~\ref{fig:tL2}). 
 
As an example, we show the $M_{\rm min}$ trajectory for iPTF14hls ($t_{\rm pl} \sim10^3\,\rm days$; $L_{\rm pl} \simeq3\times10^{42}\,\rm erg\,s^{-1}$) as a red dashed line in the bottom panel of Fig.~\ref{fig:tL2} (along this line, acceleration of the ejecta is negligible). If the velocity obtained from the Fe line absorption minimum, $v\simeq4000\,\rm km\,s^{-1}$ \citep{Arcavi+2017d}, is representative of the unshocked ejecta, then we find the initial explosion energy and ejecta mass are constrained to be $E_0\simeq10^{52}\,\rm erg$ and $M_{\rm ej} \gtrsim M_{\rm min}\simeq200\,\Msun$, respectively. Such a large ejecta mass would obviously require a very massive progenitor star, possibly supporting a pulsational pair-instability explanation for this event \citep[e.g.,][]{Woosley+2007,Woosley2018,Woosley&Smith22,Wang+2022,Chen+23}. 

\begin{figure}
\begin{center}
\includegraphics[width=85mm, angle=0,bb=0 0 323 216]{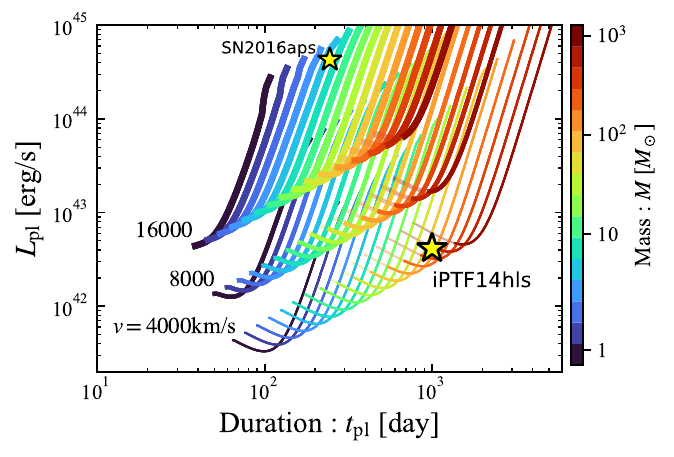}
\includegraphics[width=85mm, angle=0,bb=0 0 273 216]{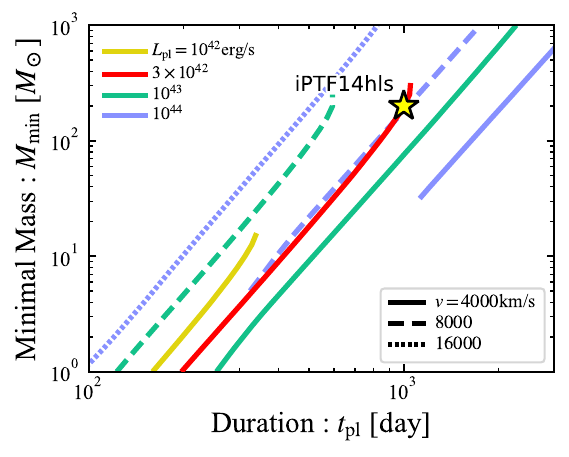}
\caption{({\bf Top}) The same as the top panel of Fig.~\ref{fig:tL2} but for fixed $v$. The trajectories terminates at high luminosity (roughly corresponding to $H_{\rm pl,max}$, where the ejecta accelerates). ({\bf Bottom}) Minimal ejecta mass $M_{\rm min}$ to produce a given plateau duration $t_{\rm pl}$ for different luminosity and velocity.}
\label{fig:tL3}
\end{center}
\end{figure}

The minimum ejecta mass (Eq.~\eqref{eq:Mminimal}) can be expressed in terms of $v$ instead of $E_0$:
\begin{eqnarray}
M_{\rm min}&\simeq& 140\,\Msun\,\left(\frac{L_{\rm pl}}{3\times10^{42}{\,\rm erg\,s^{-1}}}\right)^{-1/2}\nonumber \\
&&\left(\frac{v}{4000{\,\rm km\,s^{-1}}}\right)^{3}\left(\frac{t_{\rm pl}}{10^3{\,\rm day}}\right)^{3}\ ,
    \label{eq:Mminimal2}
\end{eqnarray}
where again acceleration of the ejecta is neglected.  The bottom panel of Fig.~\ref{fig:tL3} shows $M_{\rm min}$ as a function of plateau duration for different values of $L_{\rm pl}$ and $v$. Each $M_{\rm min}$ curve terminates at both short and long plateau durations, the latter corresponding to the regime of appreciable acceleration of the ejecta ($H\gtrsim H_{\rm pl,max}$). This is illustrated in the top panel of Fig.~\ref{fig:tL3}, which is similar to Fig.~\ref{fig:tL2} but shows $L_{\rm pl}-t_{\rm pl}$ trajectories for different assumed mean ejecta velocities.

The above demonstrates a methodology for placing lower limits on the ejecta mass for given supernovae. This can help identify the explosions of very massive stars, including PISNe caused by pair-production-induced collapse leading to explosive oxygen burning for progenitors with $\simeq140-260\,\Msun$ \citep{Rakavy&Shaviv1967,Barkat+1967,Heger&Woosley2002,Heger+2003,Kasen+2011,Dessart+2013,ChenKeJung+2014c,Kozyreva+2014,Kozyreva+2014b,Gilmer+2017} and general-relativistic instability supernovae (GRSNe) for supermassive population III stars $\sim10^{4-5}\,\Msun$ \citep{Chandrasekhar1964,Fuller+1986,Montero+2012,ChenKeJung+2014,Matsumoto+2016,Uchida+2017,Moriya+2021b,Nagele+2020,Nagele+2022}. These explosions should be rare and current observations only weakly constrain their rates. For instance, \citet{Moriya+2021} constrain the rate of long-lasting H-rich supernovae observationally, finding $\sim 40\%$ of type IIn exhibit light-curves lasting longer than a year, while the rate of PISN is constrained to $\lesssim 0.01-0.1\%$ of the total core-collapse supernova rate. More recently, using \textit{JWST} observations, \cite{Moriya+2023c} constrain the rate of GRSNe to $\lesssim800\,\rm Gpc^{-3}\,yr^{-1}$. With current and upcoming facilities such as \textit{JWST}, \textit{Rubin}, \textit{Euclid}, and \textit{Roman} in an excellent position to discover additional candidate explosions in the near future, our framework may prove useful to help identify these events and constrain the ejecta properties.

\subsection{Application to Luminous Red Novae}
Luminous red novae (LRNe) are weeks- to months-long transients with luminosities between those of classical novae and supernovae (e.g., \citealt{Bond+2003,Tylenda+2011,Blagorodnova+2017}), and which are believed to arise from stellar mergers \citep{Tylenda&Soker2006,Ivanova+2013b}. While dimmer members of the LRN class with $\lesssim10^{40}\,\rm erg\,s^{-1}$ can be powered exclusively by hydrogen recombination energy, the brightest and longest events may require an additional energy source \citep{Matsumoto&Metzger2022b}, most likely shock interaction between the merger ejecta and stellar material unbound from the binary prior to the dynamical merger \citep[e.g.,][]{Pejcha+2016,Pejcha+2016b,Metzger&Pejcha2017,MacLeod&Loeb2020}.

 Our results illustrate that shock interaction can increase the plateau duration and luminosity by at most a factor $\sim 2$ and $\sim10$, respectively, boosting the radiated energy by a factor $\lesssim 20$ compared to the zero-heating case. Scaling to properties typical of stellar mergers, the duration, luminosity, and radiated energy in the zero-heating case are estimated by Eqs.~\eqref{eq:tpl_popov}, \eqref{eq:Lpl_popov}, and \eqref{eq:Epl_popov}:\footnote{Note that for these parameters, the ejecta energy,
\begin{align}
E_0\simeq9.0\times10^{48}{\,\rm erg\,}\left(\frac{M}{3\,\Msun}\right)\left(\frac{v}{10^3\,{\rm km\,s^{-1}}}\right)^2\ ,
    \nonumber
\end{align}
is close to the limit below which gas pressure is comparable to radiation pressure (Eq.~\eqref{eq:Prad>Pgas}).}
\begin{align}
t_{\rm pl,Popov}&\simeq78{\,\rm day\,}\left(\frac{M}{3\,\Msun}\right)^{5/14}
    \nonumber\\
&\left(\frac{v}{10^3\,{\rm km\,s^{-1}}}\right)^{-5/14}\left(\frac{R_0}{30\,\Rsun}\right)^{1/7}\ ,\\
L_{\rm pl,Popov}&\simeq2.6\times10^{40}{\,\rm erg\,s^{-1}\,}\left(\frac{M}{3\,\Msun}\right)^{3/7}
    \nonumber\\
&\left(\frac{v}{10^3\,{\rm km\,s^{-1}}}\right)^{11/7}\left(\frac{R_0}{30\,\Rsun}\right)^{4/7}\ ,\\
E_{\rm pl,Popov}&\simeq1.7\times10^{47}{\,\rm erg\,}\left(\frac{M}{3\,\Msun}\right)^{11/14}
    \nonumber\\
&\left(\frac{v}{10^3\,{\rm km\,s^{-1}}}\right)^{17/14}\left(\frac{R_0}{30\,\Rsun}\right)^{5/7}\ .
\end{align} 
While these estimated fall short of explaining the most luminous LRNe by an order of magnitude for reasonable ejecta masses \citep{Matsumoto&Metzger2022b}, the presence of CSM interaction could alleviate these tensions by boosting the plateau luminosity and duration for optimally-chosen values of the (pre-merger) binary mass-loss rate, $\dot{M} \sim \dot{M}_{\rm cr}$ (Sec.~\ref{sec:shock interaction}). In particular, from Eq.~\eqref{eq:Mdot_cr} we find:
\begin{align}
\dot{M}_{\rm cr}&\simeq1.34{\,\Msun\,\rm yr^{-1}\,}\left(\frac{M}{3\,\Msun}\right)
    \nonumber\\
&\left(\frac{v}{10^3\,{\rm km\,s^{-1}}}\right)^{-2}v_{\rm CSM,100}\left(\frac{\zeta}{0.56}\right)^{-3}\ ,
\end{align}
where here we take $\zeta \equiv v_{\rm sh}/v \simeq 0.56$, appropriate for LRN parameters. Since mass-loss prior to the merger (e.g., from the binary L2 Lagrange point) typically amounts to at most tens of percent of the donor star's mass \citep{MacLeod+2017,Pejcha+2017,MacLeod&Loeb2020b}, this optimal mass-loss rate could be realized in a massive $\sim10\,\Msun$ binary which loses $\sim M_{\odot}$ over the year preceding the dynamical merger event.

\section{Summary and Conclusions}
\label{sec:conclusions}

The basic observables of SNeIIP contain information on the properties of the exploding star, which must be understood to take advantage of the growing samples of supernovae to be discovered, e.g. by {\it Rubin}, {\it JWST}, and {\it Roman}. One such readily-measured property is the duration of the optically-thick plateau phase, which canonically depends (for a given explosion energy and progenitor radius) on the ejecta mass. All else being equal, a longer plateau requires a higher ejecta mass, and hence more massive progenitor star. However, the plateau duration can also be prolonged by the presence of a sustained internal heating source, which keeps the ejecta ionized$-$and hence optically-thick$-$longer than in a passively cooling case. Indeed, growing evidence points to additional heating sources being required to power a large number of peculiar supernovae and related optical transients. 

To understand the interplay between these processes, and how to break degeneracies to better constrain the ejecta properties of individual events, we have studied the effects of an internal heating source on light curves of hydrogen-rich explosions by generalizing the analytical SNIIP light curve model of \cite{Popov1993}. Our findings are summarized as follows:

\begin{itemize}
\item While sufficiently large internal heating rates can boost the plateau luminosity by several orders of magnitude, the plateau duration can be prolonged by at most a factor of $\sim2-3$ times compared to zero-heating case. For a temporally-constant heating source, the maximal plateau duration is realized for the maximal heating rate which does not appreciably alter the ejecta dynamics. Above this critical heating rate, the ejecta experiences significant acceleration, reducing the photon diffusion timescale and hence shortening the plateau. 
\item Heating from $^{56}$Ni decay is found to boost the plateau duration by at most a few tens of percent for nickel masses typical of SNIIP, consistent with previous findings in the literature. This agreement with more sophisticated radiation transport calculations supports our model capturing at least semi-quantitatively the effects of an internal heating source on supernova light-curves.
\item Insofar that sustained energy injection from a central engine, such as millisecond magnetar or accreting compact object, can be described as heating with a characteristic duration and total deposited energy as in Eq.~\eqref{eq:H(t)_NS} or \eqref{eq:H(t)_BH}, the resulting supernova plateau phase can exhibit a wide range of luminosities and durations depending on these parameters.  While a longer engine heating timescale (e.g., magnetar spin-down time $t_{\rm sd}$, or the mass fall-back timescale $t_{\rm fb}$) results in a longer plateau duration, the latter is again capped at a few times the duration in the zero-heating case in the limit of temporally-constant heating ($t_{\rm sd}, t_{\rm fb} \rightarrow \infty$).
\item Shock interaction between the fast supernova ejecta and slower equatorially-confined (``disk-like'') CSM can provide an additional source of sub-photospheric heating for sustaining a longer plateau (Fig.~\ref{fig:picture}). However, for typical parameters, shock heating can boost the plateau duration by at most a factor of $\sim 2$ for an optimal mass loss rate (Eq.~\eqref{eq:Mdot_cr}). While a very massive CSM results in a high shock luminosity, concomitant strong deceleration of the ejecta limits this phase to early times. By contrast, while ejecta deceleration is weaker for low CSM masses, in that case the shocks quickly expand beyond the retreating photosphere radius, therefore no longer serving as a sustained {\it internal} heating source capable of keeping the ejecta ionized.

\end{itemize}
Our results are conveniently summarized in the space of plateau duration and luminosity, as shown in Fig.~\ref{fig:tL} and compared to observed hydrogen-rich supernovae. 
\begin{itemize}
\item Motivated by our finding that the temporally-constant heating case defines an absolute boundary on the attainable duration and luminosity, we developed a framework to constrain the minimal mass for observed events, as captured in Eq.~\eqref{eq:Mminimal2} in Sec.~\ref{sec:min_mass}. As a proof of principle, we used the observed duration, luminosity, and ejecta velocity, to constrain the ejecta mass in iPTF14hls to $\gtrsim200\,\Msun$, supporting an explosion of very massive star, such as a PISN.
\item Our results can also be applied to shock interaction heating in LRNe from stellar mergers. We confirmed that shock interaction between the LRN ejecta and circumbinary material ejected prior to the merger can indeed explain the otherwise puzzling properties of the most luminous events, for an assumed pre-dynamical binary mass-loss rate of $\sim\Msun\,\rm yr^{-1}$.
\end{itemize}

\begin{acknowledgements}
T.M. acknowledges supports from JSPS Overseas Research Fellowship, the Hakubi project at Kyoto University, and JSPS KAKENHI (grant number 24K17088).  B.D.M. acknowledges support from the National Science Foundation (grant number AST-2009255). The Flatiron Institute is supported by the Simons Foundation.    
\end{acknowledgements}

\appendix
\section{Validation of the one-zone model}
\label{asec:validation}

Here we discuss the validity of the one-zone model employed in our study. As briefly discussed in Sec.~\ref{sec:method}, the original works by \cite{Arnett1980,Popov1993} did not take the one-zone approach but derived the radial distribution of temperature by solving the diffusion equation. We nevertheless found their light curve predictions to be reasonably reproduced by the one-zone model. This may partly result from the fact that the \cite{Arnett1980} temperature profile is relatively flat, being described by a Bessel-type function, $T^4 \propto \sin(r)/r$.

We first reconsider the original model by \cite{Arnett1980,Popov1993}. One of the key assumptions of their solution is that the temperature profile exhibits a self-similar profile, which is not always a good approximation particularly at early times. A self-similar profile is established not instantaneously but instead gradually over a timescale of $\sim t_{\rm diff}$ as can be seen in simulations by e.g., \cite{Khatami&Kasen2019}. However, in the case of SNIIP, the (initial) temperature profile is set not by radiative diffusion from the inner region, but by heating from the supernova shock. As shown in previous studies \citep[e.g.,][]{Bersten+2011,Goldberg+2019}, the resulting radial profile exhibits a relatively flat shape due to the nature of the shock heating, which could be nicely approximated by the self-similar solution of \cite{Arnett1980} and the one-zone model. Furthermore, this flat temperature profile does not evolve significantly with time, its normalization just being reduced by adiabatic cooling, until a diffusion wave or recombination front reaches each layer.

The situation becomes more complex when a heating source is present within the ejecta. As discussed in Sec.~\ref{sec:physical}, this heating may be spatially concentrated, which challenges the one-zone approach. An ultimate validation will require detailed radiation-hydrodynamical simulations, but there are several reasons that the one-zone model employed in this study could still be justified to first order . In SNIIP, the diffusion timescale is already comparable to the plateau duration without heating (see Eqs.~\ref{eq:tdiff} and \ref{eq:tpl_popov}) for typical parameter values. Since our main focus is on quantifying how the plateau duration can be extended by a heating source, the diffusion timescale is thus almost always shorter than the (lengthened) plateau duration. Thus the effects of radiative diffusion are likely to begin already during the recombination phase, bringing most of the ejecta into causal contact and thus justifying a single-zone approach. In cases when the external energy source is strong enough to affect the ejecta dynamics, the one-zone description becomes more applicable. This is because, while in our model we treat the ejecta acceleration as being driven exclusively by internal pressure (see Eq.~\ref{eq:EoM}), a portion of this acceleration will occur as the result of a shock for dynamically relevant heating rates. In particular, if radiative diffusion cannot transport the energy being deposited by the external source, it will accumulate around the location it is deposited, eventually driving a shock wave, which propagates outwards, both heating and accelerating the ejecta. Similar to the above argument regarding normal SNIIP (heated by the initial supernova shock), such an external heating-induced shock will also establish a relatively flat temperature profile, which again can be described reasonably well by the one-zone model. The shock-driven acceleration may furthermore trigger the Rayleigh-Taylor instability \citep[e.g., see][]{Suzuki&Maeda2021}, which mixes and homogenizes the ejecta.

The above arguments hold only for SNIIP and less massive transients such as LRNe. However, for explosions with more massive ejecta, the diffusion timescale can be longer than the plateau duration obtained by our model, invalidating the one-zone approach. For a given heating luminosity, the critical mass above which the diffusion time becomes longer than the plateau duration is given by comparing Eqs.~\eqref{eq:tdiff} and \eqref{eq:tpl}:
\begin{align}
M_{\times}\simeq8300\,\Msun\,v_{6000}^{-3}T_{\rm i,6000}^{-4}H_{43}\ .
    \label{eq:M_violation}
\end{align}
While large compared to typical stellar masses, this limit should be kept in mind when carrying out broad parameter studies. In fact, for our calculations presented in Sec.~\ref{sec:discussion}, this condition is violated for the highest ejecta mass cases, which we therefore mark with transparent lines in Figs.~\ref{fig:tL2} and \ref{fig:tL3}.

\section{Analytic Estimates for Broken Power-law Heating Profile}
\label{sec:powerlaw}

We now extend our analytic estimates from Sec.~\ref{sec:constant} for a constant heating rate to the broken power-law heating rate of the form \eqref{eq:H(t)}. We first consider cases in which the heating affects only the thermal evolution of ejecta as in Sec.~\ref{sec:constant_v}. A quantitatively different evolution from the $H=const$ case occurs for $\tbr<t<\ti$. In this regime, as long as $\alpha<2$ the internal energy grows as $E\simeq tH\propto t^{1-\alpha}$ after the deposited energy has become comparable to the internal energy at time
\begin{align}
\widetilde{t}_{\rm h}=\left(\frac{t_{\rm h}}{\tbr}\right)^{\frac{\alpha}{2-\alpha}}t_{\rm h}\ ,
    \label{eq:th_tilde}
\end{align}
where the definition for $t_{\rm h}$ follows Eq.~\eqref{eq:th} but now for $H\to \widetilde{H}$. The evolution of the internal energy and effective temperature subsequently obey:
\begin{align}
E&\simeq \left(\frac{t}{\tbr}\right)^{1-\alpha}\widetilde{H}\tbr \ ,
	\label{eq:E(t)_2}\\
T_{\rm eff}&\simeq \left(\frac{\widetilde{H}}{H_{\rm cr}}\right)^{1/4}\left(\frac{t}{\tbr}\right)^{-\alpha/4}T_{\rm i} \ ,
    \label{eq:Teff3_h}
\end{align}
respectively. For sufficiently high heating rates, $\widetilde{H}>H_{\rm cr}$, recombination begins at time
\begin{align}
\widetilde{t}_{\rm i}=\left(\frac{\widetilde{H}}{H_{\rm cr}}\right)^{1/\alpha}\tbr\ .
    \label{eq:ti2}
\end{align}
When $\alpha\geq2$, the evolution until recombination completes is identical to the zero-heating case.  

From Eq.~\eqref{eq:th_tilde} we find that for heating rates smaller than a critical rate, $\widetilde{H}<t_0E_0/\tbr^2$, the thermal evolution of the ejecta is affected only after the heating rate has begun to decline ($\widetilde{t}_{\rm h}>\tbr$). For higher heating rates, the evolution is modified before the heating breaks ($t_{\rm h}<\tbr$), and for $\alpha<2$ it is identical to the $H=const$ case until $t<\tbr$. The evolution at $t>\tbr$ is identical to that given in Eqs.~\eqref{eq:E(t)_2}, \eqref{eq:Teff3_h}. For $\alpha\geq2$, the internal energy is effectively reset by the injected heat on the time $\tbr$, such that $E\simeq \widetilde{H}\tbr(t/\tbr)^{-1}$. And the effective temperature and recombination time follow the same expressions as in Eqs.~\eqref{eq:Teff1_h} and \eqref{eq:ti} for the regime of $\tdiff<t$ and $H_{\rm cr}<H$, respectively, but making the replacements $\tdiff\to\tbr$ and $H\to \widetilde{H}$.

The photosphere evolution during the recombination phase can be derived following similar arguments as in the constant heating case.  In particular, when the heating rate declines before recombination begins ($\tbr<\ti$), the photosphere location follows
\begin{align}
x\simeq\begin{cases}
\left(\frac{t}{t_{\rm i}}\right)^{-2/5}&:t<\widetilde{t}_{\rm h}^{\rm rec}\ ,\\
\left(\frac{\widetilde{H}}{2H_{\rm cr}}\right)^{1/4}\left(\frac{t}{\tbr}\right)^{-\alpha/4}&:\widetilde{t}_{\rm h}^{\rm rec}<t<\widetilde{t}_{\rm diff}^{\rm rec}\ ,\\
\left(\frac{\widetilde{H}}{H_{\rm cr}}\right)^{1/2}\left(\frac{\tdiff}{\tbr}\right)\left(\frac{t}{\tbr}\right)^{-\frac{\alpha+2}{2}}&:\widetilde{t}_{\rm diff}^{\rm rec}<t\ ,
\end{cases}
\end{align}
where 
\begin{align}
\widetilde{t}_{\rm h}^{\rm rec}&=\left(\frac{\widetilde{H}}{2H_{\rm cr}}\right)^{\frac{5}{5\alpha-8}}\left(\frac{\tbr}{\ti}\right)^{\frac{5\alpha}{5\alpha-8}}\ti\ ,\\
\widetilde{t}_{\rm diff}^{\rm rec}&=\left(\frac{2\widetilde{H}}{H_{\rm cr}}\right)^{\frac{1}{\alpha+4}}\left(\frac{\tbr}{\tdiff}\right)^{\frac{\alpha}{\alpha+4}}\tdiff\ .
	\label{eq:trec2_2}
\end{align}
Depending on the parameters, some regimes disappear; for instance, when the break in the heating rate occurs after recombination completes ($\tbr>\ti$), the evolution follows Eq.~\eqref{eq:x(t)} until the break. However, for any set of parameters, the final regime of evolution always exists, with the condition $x\tau=1$ again giving the plateau duration:
\begin{align}
\widetilde{t}_{\rm pl}\simeq\left(\frac{\widetilde{H}}{H_{\rm cr}}\right)^{\frac{1}{\alpha+6}}\left(\frac
{v}{c}\right)^{\frac{1}{\alpha+6}}\left(\frac{\tbr}{t_{\rm thin}}\right)^{\frac{\alpha}{\alpha+6}}t_{\rm thin}\ .
    \label{eq:tpl4}
\end{align}
As expected, all of the expressions given here except for Eq.~\eqref{eq:ti2} reduce to the $H=const$ case in the limit $\alpha\to0$.

We now discuss cases in which the heating impacts the ejecta dynamics, as considered in Sec.~\ref{sec:evolving_v}. For such large heating rates, the ejecta experiences appreciable acceleration at times $t>t_{\rm acc}$, where $t_{\rm acc}$ is again defined by Eq.~\eqref{eq:tacc} but for $\widetilde{H}$ instead of $H$. While in the $H=const$ case the acceleration was found to terminate when recombination occurs or photon diffusion becomes important, we find that acceleration can also terminate after the break in the heating-rate. This is because the internal energy falls as $E\propto t^{1-\alpha}$ ($\alpha<2$) or $\propto t^{-1}$ ($\alpha\geq2$), which results in an equation of motion: $dv/dt\propto1/(t^\alpha v)$ or $\propto1/(t^2 v)$, respectively.  Therefore, as long as $\alpha>1$, as obeyed by the physically-motivated heating rates we consider, the ejecta is no longer accelerated significantly after the break. This simplifies the analysis by allowing us to treat the acceleration in the same way as in the constant-heating case. In particular, we need only consider two cases, in which the ejecta acceleration terminates due to the break at $\tbr$ or photon diffusion at $\tdiff$. In the former case, the terminal velocity is given by
\begin{align}
\widetilde{v}_{\rm max}\simeq\left(\frac{10\widetilde{H}\tbr}{3M}\right)^{1/2}\ ,
\end{align}
and the corresponding photon diffusion timescale is given by
\begin{align}
\widetilde{t}_{\rm diff}^{\rm acc}&=\left(\frac{27\kappa^2M^3}{160\pi^2c^2\tbr \widetilde{H}}\right)^{1/4}\ .
\end{align}
Therefore, the condition $\tbr<\widetilde{t}_{\rm diff}^{\rm acc}$ is satisfied for
\begin{align}
\widetilde{H}<\frac{27\kappa^2M^3}{160\pi^2c^2\tbr^5}\ .
\end{align}
The plateau duration is obtained in the same way as the $H=const$ case, namely, following Eq.~\eqref{eq:tpl4}/\eqref{eq:tthin} with $\widetilde{v}_{\rm max}$ replacing $v_{\rm max}$ when recombination occurs after/before $\tbr$. In cases of yet stronger heating, acceleration terminates only once photon diffusion becomes important. The corresponding velocity and plateau duration in this case again follow the Eqs.~\eqref{eq:v2} and \eqref{eq:tpl^acc}, respectively. 

\section{Details of the ejecta-CSM shock dynamics}\label{sec:shock2}
We describe our treatment of the dynamics of the shock interaction between the supernova ejecta and equatorially-concentrated (disk-like) CSM for the case of a generic power-law density profile \citep[see also][]{Moriya+2013d,Metzger&Pejcha2017,Hiramatsu+2024}. We assume that both the forward shock (FS) and reverse shock (RS) are radiative and hence both share a single representative radius $R_{\rm sh}$ and (lab frame) velocity $v_{\rm sh}$. The unshocked CSM is characterized by its solid angle $4\pi f_\Omega$, velocity $v_{\rm CSM} \ll v_{\rm sh}$, and a radial density profile
\begin{align}
\rho_{\rm CSM}=Ar^{-q}\ ,
\end{align}
respectively. The mass $M_{\rm sh}$ and momentum $M_{\rm sh}v_{\rm sh}$ of the shocked gas increase as the FS and RS sweep up the CSM and supernova ejecta, respectively, their evolution obeying:
\begin{align}
\frac{dM_{\rm sh}}{dt}&=4\pi f_\Omega R_{\rm sh}^2\rho_{\rm SN}\left(v_{\rm SN}-v_{\rm sh}\right)+4\pi f_{\Omega} R_{\rm sh}^{2-q}A(v_{\rm sh}-v_{\rm CSM})\ ,
    \label{eq:shock_mass}\\
\frac{d(M_{\rm sh}v_{\rm sh})}{dt}&=4\pi f_\Omega R_{\rm sh}^2\rho_{\rm SN}v_{\rm SN}\left(v_{\rm SN}-v_{\rm sh}\right)+4\pi f_{\Omega} R_{\rm sh}^{2-q}Av_{\rm CSM}(v_{\rm sh}-v_{\rm CSM})\ ,
    \label{eq:shock_mom}
\end{align}
where $\rho_{\rm SN}=M/(4\pi R^2/3)$ is the ejecta density (taken to be a homogeneous sphere of radius $R$) and $v_{\rm SN}=v(R_{\rm sh}/R)$ its velocity (assuming homologous expansion). Momentum conservation (Eq.~\ref{eq:shock_mom}) can be rewritten:
\begin{align}
M_{\rm sh}\frac{dv_{\rm sh}}{dt}&=4\pi f_{\Omega}R_{\rm sh}^2\rho_{\rm SN}(v_{\rm SN}-v_{\rm sh})^2-4\pi f_{\Omega} R_{\rm sh}^{2-q}A(v_{\rm sh}-v_{\rm CSM})^2\ .
    \label{eq:shock_eom}
\end{align}
The dynamics of the swept-up shell is thus determined by solving Eqs.~\eqref{eq:shock_mass} and \eqref{eq:shock_eom}  with initial conditions $R_{\rm sh}=R_0$, $v_{\rm sh}=0$, and $M_{\rm sh}\ll M$ for assumed values of $A$, $q$, $v_{\rm CSM}$, and $f_{\Omega}$.

Once the shock dynamics are determined, we estimate the total shock luminosity (ejecta heating rate) as the sum of the kinetic luminosities of the FS and RS shocks, given, respectively, by:
\begin{align}
L_{\rm kin,FS}&=4\pi f_{\Omega}R_{\rm sh}^2(v_{\rm sh}-v_{\rm CSM})e_{\rm FS}=\frac{9\pi}{2}f_{\Omega}R_{\rm sh}^{2-q}A(v_{\rm sh}-v_{\rm CSM})^3\ ,\\
L_{\rm kin,RS}&=4\pi f_{\Omega}R_{\rm sh}^2(v_{\rm SN}-v_{\rm sh})e_{\rm RS}=\frac{27f_\Omega MR_{\rm sh}^2(v_{\rm SN}-v_{\rm sh})^3}{8R^3}\ ,
\end{align}
where the internal energy density of the post FS (RS) region is given by
\begin{align}
e_{\rm FS(RS)}&=\frac{2\rho_{\rm CSM(SN)}(v_{\rm sh}-v_{\rm CSM(SN)})^2}{(\gamma+1)(\gamma-1)}\ .
\end{align}
Here $\gamma$ is the adiabatic index and we adopt $\gamma=5/3$. In a more detailed calculation, the dynamics of the shocks would be coupled to that of the supernova ejecta via heating, such that their evolution should be calculated self-consistently.

\begin{figure}
\begin{center}
\includegraphics[width=85mm, angle=0,bb=0 0 284 230]{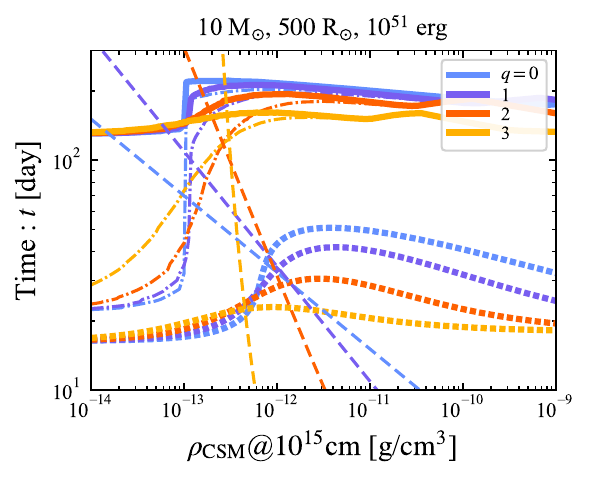}
\caption{The same as Fig.~\ref{fig:time_shock} but for different slopes of CSM radial density profile, $\rho_{\rm CSM}\propto r^{-q}$ (indicated by different colors).  Thick solid lines denote the plateau duration and dotted lines denote the time at which $T_{\rm eff}=T_{\rm i}$. Dashed diagonal lines denote the shock deceleration timescale. The horizontal axis denotes the CSM density at $10^{15}\,\rm cm$.} 
\label{fig:time_shock_v2}
\end{center}
\end{figure}

Fig.~\ref{fig:time_shock_v2} shows the key timescales plot (cf.~Fig.~\ref{fig:time_shock}) as a function of the CSM density normalization at $r=10^{15}\,\rm cm$ for different values of the CSM power-law slope $q$ as marked. The qualitative features in all cases follow those in the standard $q=2$ wind-like profile. The impact on the recombination time (thick dashed curves) and the plateau duration (thick solid curves) are slightly larger for smaller $q$. This may be because for $q<2$, the shock luminosity increases with time, peaking on the deceleration time.\footnote{An increasing heating rate potentially pushes the photosphere outward during the recombination phase, $dx/dt>0$. This occures when the dimensionless phoptosphere becomes smaller than 
\begin{align}
x_{\rm cr}=\frac{1}{2}\left\{-\left(\frac{t}{t_{\rm diff}}\right)^2+\sqrt{\left(\frac{t}{t_{\rm diff}}\right)^4+8\left(\frac{H}{H_{\rm cr}}\right)}\right\}^{1/2}\ ,
\end{align}
while we confirm that the photophere always shrinks in our calculations.} Furthermore, the fact that deceleration starts earlier helps to satisfy the condition \eqref{eq:Rsh<Rph} for a longer time. Nonetheless, the plateau duration is not extended more than a few times the zero-heating case ($\rho_{\rm CSM} \rightarrow 0$), even for $q \ne 2.$

\section{Glossary of symbols and notations}
Table~\ref{table:notation} summarizes the symbols and notations of characteristic quantities.

\begin{table*}
\begin{center}
\caption{Definitions of Key Timescales and Heating Rates}
\label{table:notation}
\begin{tabular}{lcl}
\hline\hline
Notation&Eq.&Meaning\\
\hline\hline
$t_{\rm dyn}$&\eqref{eq:tdyn}&Dynamical time\\
$\tdiff$&\eqref{eq:tdiff}&Diffusion time \\
$\ti$&\eqref{eq:ti}&Recombination time (beginning of recombination phase)\\
$t_0$&\eqref{eq:t0}&Initial dynamical time\\
$\th$&\eqref{eq:th}&Heating timescale (heating starts to modify thermal evolution)\\
$t_{\rm h}^{\rm rec}$&\eqref{eq:th^rec}&Heating timescale during recombination phase\\
$t_{\rm diff}^{\rm rec}$&\eqref{eq:tdiff^rec}&Diffusion timescale during recombination phase (after which $L\simeq H$)\\
$t_{\rm pl}$&\eqref{eq:tpl}&Plateau duration (when ejecta becomes optically thin) including heating effects\\
$t_{\rm thin}$&\eqref{eq:tthin}& When fully ionized ejecta becomes optically thin\\
$t_{\rm acc}$&\eqref{eq:tacc}&Acceleration timescale (heating source starts to significantly accelerate ejecta)\\
$t_{\rm diff}^{\rm acc}$&\eqref{eq:tdiff^acc}&Diffusion timescale for accelerated ejecta\\
$t_{\rm i}^{\rm acc}$&\eqref{eq:ti^acc}&Recombination timescale for accelerated ejecta\\
$t_{\rm pl}^{\rm acc}$&\eqref{eq:tpl^acc}&Plateau duration for accelerated ejecta\\
$t_{\rm pl,max}$&\eqref{eq:tpl_max}/\eqref{eq:tpl_max2}&Maximal plateau duration (\eqref{eq:tpl_max2} if Eq.~\eqref{eq:condition_tpl} is satisfied)\\
$t_{\rm pl,Popov}$&\eqref{eq:tpl_popov}&Plateau duration in heating-free case (Popov formulae)\\
$t_{\rm thin}^{\rm acc}$&\eqref{eq:tthin^acc}&Accelerated fully ionized ejecta becomes optically thin\\
$\tbr$ & \eqref{eq:H(t)} & Break time in broken power-law heating rate.\\ 
\hline
$H_{\rm cr}$&\eqref{eq:Hcr}&Critical heating rate (above which recombination is delayed significantly)\\
$H_{\rm min}$&\eqref{eq:Hmin}&Minimal heating rate which can apprecisely prolong plateau duration $H =0$ limit\\
$H_{\rm pl,max}$&\eqref{eq:Hpl_max}&Heating rate which maximizes plateau duration (Eq.~\eqref{eq:Hcr} if Eq.~\eqref{eq:condition_tpl} is satisfied)\\
$H_{\rm thin}$&\eqref{eq:Hthin}&Minimal heating rate required to keep ejecta ionized until it becomes optically-thin\\
\hline\hline
\end{tabular}
\end{center}
\end{table*}

\bibliographystyle{aasjournal}
\bibliography{reference_matsumoto,refs_BDM}

\end{document}